\documentclass[11pt]{article}
\usepackage[top=1in,left=1in,right=1in,bottom=1in]{geometry}

\newfont{\mycrnotice}{ptmr8t at 7pt}
\newfont{\myconfname}{ptmri8t at 7pt}
\let \originalleft \left
\let\originalright\right
\renewcommand{\left}{\mathopen{}\mathclose\bgroup\originalleft}
\renewcommand{\right}{\aftergroup\egroup\originalright}

\usepackage[usenames,dvipsnames,svgnames,table]{xcolor}
\usepackage{bm}

\newcommand{\kanat}[1]{{\color{purple}{\bf Kanat:} #1}}

\newcommand{\yihan}[1]{{\color{magenta}{\bf Yihan:} #1}}

\newcommand{\hide}[1]{} 

\newcommand{\mf}[1]{{\mbox{\sc{#1}}}}

\usepackage{amsthm}
\usepackage[ruled,lined,linesnumbered,noend]{algorithm2e}
\usepackage[noend]{algpseudocode}
\usepackage{amsmath,amssymb,subfigure,graphicx,multirow}
\usepackage{enumitem}
\usepackage{tabularx}
\usepackage{xspace}
\usepackage{alltt}
\usepackage{booktabs}
\usepackage[labelfont=bf,small,aboveskip=2pt,belowskip=2pt]{caption}            
\DeclareCaptionType{copyrightbox}

\newcommand*{\sssp}{\textsc{Sssp}\xspace}
\newcommand*{\R}{\ensuremath{\mathbb{R}}}
\newcommand*{\otilde}{\widetilde{O}\xspace}
\newcommand*{\parll}{\ensuremath{\mathbb{P}}}
\newcommand*{\polylog}{\ensuremath{\mathrm{polylog}}\xspace}
\newcommand*{\vareps}{\varepsilon}

\newcommand*{\algoName}[1]{\textsc{#1}\xspace}
\newcommand{\AlgName}{\algoName{Radius-Stepping}}
\newcommand{\ourstructure}[2]{$(#1,#2)$-ball}
\newcommand{\ourgraph}[2]{$(#1,#2)$-graph}
\newcommand{\defn}[1]{\emph{\textbf{#1}}}

\newcommand{\id}[1]{\ifmmode\mathit{#1}\else\textit{#1}\fi}
\newcommand{\const}[1]{\ifmmode\mbox{\textc{#1}}\else\textsc{#1}\fi}
\newcommand{\rk}[1]{{\bar{r}_k({#1})}}

\newtheorem{theorem}{Theorem}[section]
\newtheorem{lemma}[theorem]{Lemma}

\newtheorem{claim}[theorem]{Claim}

\newtheorem{definition}{Definition}

\begin{document}
\date{}

\title{Parallel Shortest-Paths Using Radius Stepping}
\author{Guy
  E. Blelloch\\Carnegie Mellon University\\guyb@cs.cmu.edu \and Yan Gu\\Carnegie Mellon
    University\\ yan.gu@cs.cmu.edu \and Yihan Sun\\Carnegie Mellon
    University\\ yihans@cs.cmu.edu \and Kanat Tangwongsan\\Mahidol University \\International College\\kanat.tan@mahidol.edu}

\maketitle


\begin{abstract}
%
  The single-source shortest path problem (\sssp) with nonnegative edge weights
  is a notoriously difficult problem to solve efficiently in parallel---it is
  one of the graph problems said to suffer from the transitive-closure
  bottleneck.  In practice, the $\Delta$-stepping algorithm of Meyer and Sanders
  (\emph{J. Algorithms,~2003}) often works efficiently but has no known
  theoretical bounds on general graphs.  The algorithm takes a sequence of steps,
  each increasing the radius by a user-specified value $\Delta$.  Each step
  settles the vertices in its annulus but can take $\Theta(n)$ substeps, each
  requiring $\Theta(m)$ work ($n$ vertices and $m$ edges).

  In this paper, we describe \AlgName, an algorithm with the best-known tradeoff
  between work and depth bounds for \sssp with nearly-linear ($\otilde(m)$)
  work.  The algorithm is a $\Delta$-stepping-like algorithm but uses a variable
  instead of fixed-size increase in radii, allowing us to prove a bound on the
  number of steps.  In particular, by using what we define as a vertex
  $k$-radius, each step takes at most $k+2$ substeps.  Furthermore, we define a
  $(k, \rho)$-graph property and show that if an undirected graph has this
  property, then the number of steps can be bounded by
  $O(\frac{n}{\rho} \log \rho L)$, for a total of $O(\frac{kn}{\rho} \log \rho L)$
  substeps, each parallel.  We describe how to preprocess a graph to have this
  property.  Altogether, \AlgName takes $O(m\log n)$ work
  and $O(\frac{n}{\rho}\log n \log (\rho{}L))$ depth per source after
  preprocessing.  The preprocessing step can be done in $O(m\log n + n\rho^2)$
  work and $O(\rho^2)$ depth or in $O(m\log n + n\rho^2\log n)$
  work and $O(\rho\log \rho)$ depth, and adds no more than $O(n\rho)$ edges.
\end{abstract}

\section{Introduction}
The single-source shortest path problem (\sssp) is a fundamental graph problem
that is extremely well-studied and has numerous practical and theoretical
applications.  For a weighted graph $G = (V, E, w)$ with $n = |V|$ vertices and
$m = |E|$ edges, and a source node $s \in V$, the \sssp problem with nonnegative
edge weights is to find the shortest (i.e., minimum weight) path from $s$ to
every $v \in V$, according to the weight function $w\!: E \to \R_+$, which
assigns to every edge a real-valued nonnegative weight (``distance'').  We will
assume without loss of generality that the lightest nonzero edge has weight $1$,
i.e, $\min_{e: w(e) > 0} w(e) = 1$.  Let $L = \max_{e} w(e)$.

In the sequential setting, Dijkstra's algorithm~\cite{Dijk} solves this problem
in $O(m + n\log n)$ time using the Fibonacci heap~\cite{Fib}.  This is the best
theoretical running time for general nonnegative edge weights although faster
algorithms exist for certain special cases.  Thorup~\cite{thorup1999undirected},
for example, gives an $O(m + n)$-time algorithm for when the edge weights are
positive integers.

In the parallel setting, the holy grail of parallel \sssp is an algorithm with
Dijkstra's work bound (i.e., work-efficient) that runs in small depth.  Although
tens of algorithms for the problem have been proposed over the last several
decades, none of the existing algorithms that take the same amount of work as
Dijkstra's have polylogarithmic or even $o(n)$ depth.  This has led Karp and
Ramachandran to coin the term \emph{transitive closure
  bottleneck}~\cite{karp1991parallel}.  We survey existing algorithms most
relevant to this work in Section~\ref{sec:related-work}.

Apart from the quest for a polylogarithmic-depth, work-efficient algorithm, one
could aim for an algorithm with a high degree of parallelism ($\parll = W/D$)
that is work-efficient or nearly work-efficient.  From a performance point of
view, both factors are important because work efficiency is a prerequisite for
the algorithm to quickly gain speedups over the sequential algorithm, and the
parallelism factor $\parll$ indicates how well the algorithm will scale with
processors.

This observation has fueled the design of \sssp algorithms with substantial
parallelism that are nearly work-efficient.  Spencer~\cite{spencer1997time}
shows that BFS can be solved in $O(\frac{n}{\rho} \log^2\rho)$ depth and
$O((m+n\rho^2\log\rho)\log \rho)$ work, where $\sqrt{m/n} \leq \rho$ is a
tuneable parameter.  On a weighted graph, \sssp requires
$O(\frac{n}{\rho}\log n \log(\rho{}L))$ depth and
$O((n\rho^2\log \rho+m)\log (n\rho L))$ work, where
$\log ({\rho{}L}) \leq \rho$.  At a high-level, these algorithms use limited
path-doubling to determine the shortest-path distances to about $\rho$ vertices
in each round, requiring slightly more than $\tfrac{n}\rho$ rounds in total.

On the empirical side, the $\Delta$-stepping algorithm of Meyer and
Sanders~\cite{MS03} works well on many kinds of graphs.  The algorithm has been
analyzed for random graphs, but no theoretical guarantees are known for the
general case.  $\Delta$-stepping is a hybrid of Dijkstra's algorithm and the
Bellman-Ford algorithm.  It determines the correct distances from $s$ in
increments of $\Delta$, settling in step $i$ all the vertices whose
shortest-path distances are between $i\Delta$ and $(i+1)\Delta$.  Within each
step, the algorithm resorts to running Bellman-Ford as substeps to determine the
distances to such vertices.  Altogether, the algorithm performs up to a total of
$\frac{LK}{\Delta}$ rounds of substeps, where $K$ the shortest-path distance to
the farthest vertex from $s$.

\medskip

\noindent\textbf{Our Contributions:} In this paper, we investigate the single-source
shortest-path problem for weighted, undirected graphs.  The main result is as
follows:
\begin{theorem}\label{thm:mainthm}
  Let $G$ be an undirected, weighted graph with $n$ vertices and $m$ edges. For
  $\rho \le \sqrt{n}$, there is an algorithm, \AlgName, that after a
  preprocessing phase can solve \sssp from any source in
  $O(m\log n)$ work and
  $O(\frac{n}{\rho}\log n\log{\rho{}L})$ depth.  In addition, the preprocessing
  phase takes $O(m\log n + n\rho^2)$ work and $O(\rho^2)$ depth, or
  $O(m\log n+n\rho^2\log n)$ work and $O(\rho \log \rho)$ depth.
\end{theorem}
Compared to a standard Dijkstra's implementation, the algorithm, excluding
preprocessing, is work-efficient up to only a $\log n$ factor.  Using
$\rho = \polylog(n)$, the algorithm offers at least polylogarithmic parallelism.
Using $\rho = n^{O(\vareps)}$, the algorithm has sublinear depth.  As far as we
know, this is the best tradeoff between work and depth for the problem.


Our algorithm \AlgName (Section~\ref{sec:thealgo}) can be seen as a hybrid
variant of $\Delta$-stepping and Spencer's algorithm, though simpler.  Like
$\Delta$-stepping, it performs Dijkstra's-like steps in the outer loop, finding
and settling multiple vertices at a time using (restricted) Bellman-Ford as
substeps.  Unlike $\Delta$-stepping, however, the algorithm judiciously picks a
new step size ($\Delta$) every time.  The step size is chosen so that, like in
Spencer's algorithm, about $\rho$ new vertices are settled in every step, or
otherwise it doubles the distance explored.  Yet, unlike Spencer's algorithm,
\AlgName does not perform path doubling in every iteration, leading to a
conceptually simpler algorithm with fewer moving parts.  Ultimately, the steps
of \AlgName parallel Dijkstra's execution steps, and the only invariant
necessary is a natural generalization of Dijkstra's invariant.

We prove that \AlgName runs in $O(\tfrac{n}{\rho}\log(\rho{}L))$ steps, %
each running up to $O(k)$ substeps, provided that the input graph is a
$(k, \rho)$-graph.  This is informally defined to be a graph in which every node
can reach its $\rho$ closest vertices within $k$ hops.  We show how to implement
this algorithm in $O(km\log n)$ work and
$O(k\frac{n}{\rho}\log n \log{\rho{}L})$ depth.  For unweighted graphs, it can
be implemented in $O(k(m + n))$ work and
$O(k\frac{n}{\rho}\log \rho \log^* \rho)$ depth.

However, not all graphs are $(k,\rho)$-graphs as given, but, as will be
discussed in Section~\ref{sec:ball}, they can be made so with a preprocessing
step that runs restricted Dijkstra's to discover the $\rho$ closest vertices for
all vertices in parallel.  To make the graph a $(1, \rho)$-graph, the
preprocessing step effectively adds up to $n\rho$ shortcut edges to the input
graph.  The number of shortcut edges can be reduced if we use a larger $k$.
Although this saves some memory, \AlgName will have to do slightly more work
when computing the shortest paths.  For $k > 1$, it is less trivial to generate
a $(k, \rho)$-graph that adds only a small number of edges.  We present some
heuristics in Section~\ref{sec:ball} and empirically study their quality in
Section~\ref{sec:exp}.

Our experiments aim to study the number of edges added by preprocessing and the
number of steps taken by \AlgName{}.  This is a good proxy for measuring the
depth of the algorithm experimentally. The results confirm our theory: the
number of steps taken (that is, roughly the depth) is inversely proportional to
$\rho$ for both the weighted and unweighted cases.  They further show that with
an appropriate heuristic (such as the dynamic programming heuristic from Section
\ref{sec:heuristics}) and choice of parameters, preprocessing will add a
reasonable number of edges (at most $O(m)$), thereby increasing the total work
by at most a constant.

\section{Preliminaries}

Let $G=(V, E, w)$ be an undirected, simple, weighted graph.  We assume without
loss of generality that $G$ is a connected, simple graph (i.e., no self-loops
nor parallel edges).  For $S \subseteq V$, define
$N(S)=\cup_{u\in S}\{v\,|\,(u,v)\in E\}$ to be the \defn{neighbor set} of $S$.
During the execution of standard breadth-first search (BFS) or Dijkstra's
algorithm, the \defn{frontier} is the neighbor set of all visited vertices. Let
$d(u,v)$ denote the shortest-path distance in $G$ between two vertices $u$ and
$v$.  
The \defn{enclosed ball} of a vertex $u$ is $B(u,r)=\{v\in V\,|\,d(u,v)\le r\}$.

A \defn{shortest-path tree} rooted at vertex $u$ is a spanning tree $T$ of $G$
(subtree if $G$ is unconnected) such that the path distance from the root $u$ to
any other vertex $v\in T$ is the shortest-path distance from $v$ to $u$ in $G$.
We say a graph is \defn{unweighted} if $w(e)=1$ for all $e\in E$ (so $L=1$).

We assume the standard PRAM model that allows for concurrent reads and writes,
and analyze the performance of our parallel algorithms in terms of work and
depth~\cite{JaJa92}, where \defn{work} $W$ is equal to the total number of
operations performed and \defn{depth (span)} $D$ is equal to the longest
sequence of dependent operations.  Nonetheless, all of our algorithms can also
be implemented on machines with exclusive writes.  We say that an algorithm is
\defn{work efficient} if it performs the same amount of work as the best
sequential counterpart, up to constants.


Consider two ordered sets $A$ and $B$ stored in balanced BSTs such that $|A|\le |B|$.  Recent research work shows
that set union and set difference operations can be computed either in $O(|A|\log|B|)$ work and
$O(\log |B|)$ depth~\cite{PVW83,park2001parallel,paul1983parallel}, or in $O(|A|\log|B|/|A|)$ work and
$O(\log |A|\log |B|)$ depth~\cite{blelloch1998fast,blelloch2016join}. \hide{The first EREW PRAM algorithm
with $O(|A|\log|B|/|A|)$ work and $O(\log |B|)$ depth on set union and set difference was proposed
in \cite{katajainen1994efficient} but was later shown buggy.
It is recently fixed by an unpublished paper \cite{unpublished}. }
In this paper we assume both set union
and difference can be done in $O(|A|\log|B|)$ work and $O(\log |B|)$ depth.



Finally, we need a few definitions:

\begin{definition}[Hop Distance]
  Let $\hat{d}(u,v)$ be the number of edges on the shortest (weighted) path
  between $u$ and $v$ that uses the fewest edges.
\end{definition}

\begin{definition}[$k$-radius]
  For $u \in V$, the $k$-radius of $u$, denoted by $\bar{r}_k(u)$, is
  $\min_{v\in V,\hat{d}(u,v)>k}{d(u,v)}$---that is, the closest distance to $u$ which
  is more than $k$ hops away.
\end{definition}

\section{The Radius-Stepping Algorithm}
\label{sec:thealgo}


\hide{
\begin{table}[t]
\setlength{\extrarowheight}{.5em}
\begin{tabular}{@{}l@{ }|p{2.2cm}|p{4.8cm}}
   Notation & ~~~~~~\,{Name} & ~~~~~~~~~~~~ Description \\  \hline  
  $d_i$ & Step distance & The ``scope'' to visit vertices  \\ 
  $S_i$ & Visited set & Vertices visited in first $i$ steps \\ 
  $N(S)$ & Neighbor set & Directed neighbors of vertices in $S$ \\ 
  $N^*(S)$ & $\rho$-nearest-neighbor set  & The union of the closest $\rho$ vertices from each vertex in $S$    \\
  $w^*(u,v)$ & Graph distance & $d(u,v)$ for $v\in N^*(u)$\\
\end{tabular}
\end{table}
}

\begin{algorithm}[t]
\caption{The \AlgName{} Algorithm.}
\label{alg:sssp}
\KwIn{A graph $G=(V,E,w)$, vertex radii $r(\cdot)$, and a source node $s$.}
\KwOut{The graph distances $\delta(\cdot)$ from $s$.}
    \vspace{0.5em}
\DontPrintSemicolon
$\delta(\cdot)\leftarrow +\infty$, $\delta(s)\leftarrow 0$\label{line:init1}\\
\lForEach{$v\in N(s)$} {$\delta(v)\leftarrow w(s,v)$}
$S_0\leftarrow \{s\}$, $i\leftarrow 1$\label{line:init2}\\
\While{$|S_{i-1}|<|V|$} {\label{line:loop}
    $d_{i}\leftarrow \min_{v\in V \setminus S_{i-1}}\{\delta(v)+r(v)\}$\label{line:computeleadnode}\\
    \Repeat{no $\delta(v) \le d_i$ was updated} {\label{line:substeploopstart1}
        \ForEach{$u\in V\setminus S_{i-1} ~s.t.~ \delta(u)\leq d_i$\label{line:innerloop}} {
            \ForEach{$v\in N(u)\setminus S_{i-1}$} {
                $\delta(v)\leftarrow \min\{\delta(v), \delta(u)+w(u,v)\}$\label{line:updatealldelta}\\
            }
        }
    }\label{line:substeploopend1}
    $S_i=\{v\,|\,\delta(v)\leq d_i\}$\label{line:updateset}\\
    $i=i+1$\\\label{line:loopend}
    }
\Return {$\delta(\cdot)$}
\end{algorithm}

\begin{figure}[t]
\begin{center}
        \includegraphics[width=0.5\textwidth]{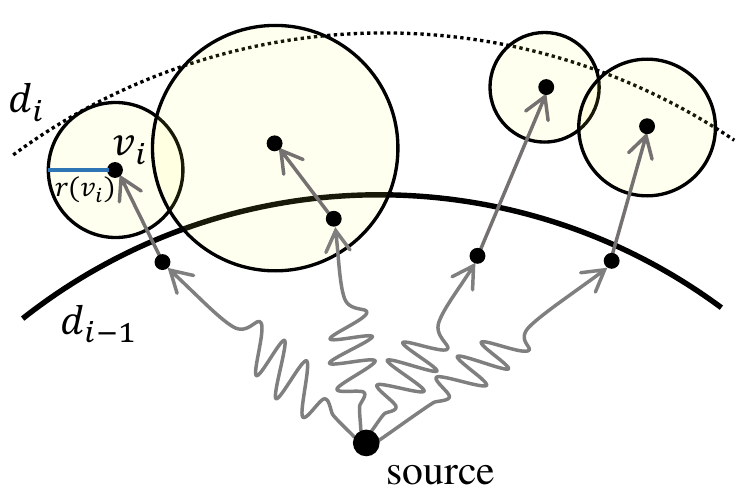}
        \caption{Illustration of a step of Algorithm~\ref{alg:sssp}: For all
          vertices whose neighbors are within $d_i$, we pick the node $v_i$, the
          lead node in the $i$-th round, that minimizes the tentative distance
          plus the vertex radius of this node.  The round distance $d_i$ is set
          to $\delta(v_i)+r(v_i)$.}\label{fig:mainalgo}
  \vspace{-2em}
\end{center}
\end{figure}

This section describes our algorithm for parallel \sssp called \AlgName{} and
discusses its performance analysis.  We begin with a high-level description,
which we analyze for the number of steps.  Following that, we give a detailed
implementation, which we analyze for work and depth.

The \AlgName{} algorithm is presented in Algorithm~\ref{alg:sssp}.  In broad
strokes, it works as follows: The input to the algorithm is a weighted,
undirected graph, a source vertex, and a target radius value for every vertex,
given as a function $r\!: V \to \R_+$.  The algorithm has the same basic
structure as the $\Delta$-stepping algorithm; both are effectively a hybrid of
Dijkstra's algorithm and the Bellman-Ford algorithm.  Like Dijkstra's algorithm,
they visit vertices in increasing distance from the source $s$, settling each
vertex $v$---i.e., determining its correct distance $d(s,v)$.  However, instead
of visiting one vertex at a time, the algorithms visit nodes in
\emph{\textbf{steps}} (the while loop in Line
\ref{line:loop}-\ref{line:loopend}). On each step $i$, the \AlgName{} increments
the radius centered at $s$ from $d_{i-1}$ to $d_i$, and settles all vertices $v$
in the annulus $d_{i-1} < d(s,v) \le d_i$.  This is illustrated in
Figure~\ref{fig:mainalgo}.

Settling these vertices involves multiple \defn{substeps} (the do loop in
Lines~\ref{line:substeploopstart1}--\ref{line:substeploopend1}) of what is
effectively Bellman-Ford's algorithm.  Each substep is easily parallelized.  In
$\Delta$-stepping, the round distance $d_i = d_{i-1} + \Delta$ increases
the radius by a fixed amount on each step.  In the worst case, this can require
$\Theta(n)$ substeps.  This is not efficient since each substep will process the
same set of vertices and their edges.  In the worst case, it could take $O(nm)$
work.

In the \AlgName{} algorithm, a new round distance $d_i$ is decided on each
round with the goal of bounding the number of substeps.  The algorithm takes a
radius $r(v)$ for each vertex and selects a $d_i$ on step $i$ by taking the
minimum of $\delta(v) + r(v)$ over all $v$ in the frontier
(Line~\ref{line:computeleadnode}).
Lines~\ref{line:substeploopstart1}--\ref{line:substeploopend1} then run the
Bellman-Ford substeps until all vertices with radius less
than $d_i$ are settled.  Vertices within $d_i$ are then added to the
visited set $S_i$.

The algorithm is correct for any radii $r(\cdot)$, but by setting it
appropriately, it can be used to control the number of substeps per
step and the number of steps.  If $r(v) = 0$, then the algorithm is
effectively Dijkstra's, and the inner step is run only once.  It may, however, visit multiple
vertices with the same distance.  If $r(v) = \infty$, then the
algorithm is effectively the Bellman-Ford algorithm, and the substeps will run
until all vertices are settled, and hence there will be a single step.
If $r(v) = \Delta$, then the algorithm is almost $\Delta$-stepping, but
not quite since $\Delta$ is added to the distance of the nearest
frontier vertex instead of to $d_{i-1}$.

In this paper, we aim to set $r(v)$ to be close, but no more,
than the $k$-radius $\bar{r}_k(v)$.  By making $r(v)$ no more than the
$k$-radius, it guarantees that the algorithm will run at most $(k+2)$
Bellman-Ford substeps.  For our theoretical bounds, we set $k=1$ so the algorithm
only does constant substeps.

In addition to bounding the number of substeps, we want to bound the
number of steps.  To do this, we can make use of properties of the
input graph.  In particular if every vertex $v$ has $\rho$ vertices
that are within a distance $r(v)$, i.e. $|B(v,r(v))|\ge \rho$, then we can
bound the number of steps by $O(\tfrac{n}{\rho} \log (\rho L))$, where $L$ is the
longest edge in the graph.  

We will now prove the bound for the number of steps and substeps, then discuss
the implementation details of Algorithm~\ref{alg:sssp} in
Section~\ref{sec:complexity}. In the analysis, we define the \emph{lead node} in
step $i$ as the node $v$ that attains $d_i$ in Line~\ref{line:computeleadnode}
and denote it by $v_i$.

\hide{
In each step, we first update the tentative distances using all
outgoing edges from visited nodes.  Then an unvisited node $v_i$ is
selected such that $v_i=\arg\min_{v\in
  N(S_{i-1})\backslash S_{i-1}}\delta(v)+r(v)$.  This node is heavily used in our
analysis, so we name it the \defn{lead node} in round $i$ and let
$d_i=\delta(v_i)+r(v_i)$.  After acquiring $d_i$ and $v_i$, for each
unvisited vertex with tentative distance no more than $d_i$, we relax
all of its $\rho$-nearest vertices.  Finally, vertices with tentative
distances no more than $d_i$ will be added to the visited set and
their distances to the source is finalized.}

\begin{theorem} [Correctness]
  \label{thm:correctness}
  After the $i$-th step of \AlgName{}, each vertex $v$ with
  $d_{i-1}<d(s,v)\le d_i$ will have $\delta(v)=d(s,v)$ and will be included in
  $S_i$.
\end{theorem}

Correctness of the algorithm is straightforward since it is essentially the
$\Delta$-stepping algorithm with a variable step size: At the end of the inner
loop, all vertices within the round distance are settled, and all of their direct
neighbors are relaxed.  Now we analyze the parallelism of this algorithm by
bounding the longest chain of dependencies (depth).

\begin{theorem} [The number of substeps]
\label{thm:numberofsubsteps}
The repeat-until loop
(Lines~\ref{line:substeploopstart1}--\ref{line:substeploopend1}) runs at most
$k+2$ times provided that $r(v)\le\bar{r}_k(v)$ for all $v\in V$.
\end{theorem}

\begin{theorem} [The number of steps]
  \label{thm:numberofrounds}
  The while loop (Lines~\ref{line:loop}--\ref{line:loopend}) requires no more
  than
  \[ \left\lceil\frac{n}{\rho}\right\rceil\left(1+\lceil \log_2\rho
    L\rceil\right)=O\left(\frac{n}{\rho}\log (\rho L)\right)
  \]
  steps provided that $|B(v,r(v))|\ge \rho$ for all $v\in V$.

\end{theorem}

We prove these theorems in
Sections~\ref{sec:substep} and~\ref{sec:step}.


\subsection{Number of Substeps}
\label{sec:substep}

Let $k$ be such that $r(v) \leq \rk{v}$ for all $v \in V$. To bound the number
of substeps, we use the fact that vertices with distances no more than the round
distance in the previous round $(d_{i-1})$ are all correctly computed, as given
by Theorem~\ref{thm:correctness}.  Observe that their neighbors are also
relaxed, since the distances to none of these vertices are updated in the last
substep of the previous step (the termination condition in
Line~\ref{line:substeploopend1}).



\begin{lemma}\label{lem:substep}
  For a vertex $v\in V$, if $d_{i-1}<d(s,v)\le d_i$, then its distance is
  correctly computed after $k+1$ substeps of the $i$-th step,
  i.e. $\delta(v)=d(s,v)$.
\end{lemma}

\begin{proof}

  Consider the shortest path from $s$ to $v$ that uses the fewest hops. Let $u$
  be the first vertex on this path with distance larger than $d_{i-1}$.  By
  Theorem~\ref{thm:correctness}, we have that $\delta(u)$ is relaxed to $d(s,u)$
  at the end of the $(i-1)$-th step.

  Now we prove that $v$ must be within $(k+1)$ hops from $u$.  Assume to the
  contrary that the hop distance $\hat{d}(u,v)>k+1$. Let $v'$ be the vertex on
  the shortest path from $s$ to $v$ which appears right before $v$, then
  $d(u,v')<d(u,v)\le d_i-\delta(u)\le r(u)\le\bar{r}_k(u)$, which contradicts
  the condition that $\hat{d}(u,v')=\hat{d}(u,v)-1>k$ (the definition of
  $k$-radius $\bar{r}_k(u)$ prevents this from happening).  Hence, we have
  $\hat{d}(u,v)\le k+1$, and the distance of $v$ can be computed within $k+1$
  substeps along the shortest path from $u$ to $v$.
\end{proof}

By Lemma~\ref{lem:substep}, all vertices with distances no more than $d_i$ can be computed within $k+1$ substeps.  We need an extra substep to relax all of their direct neighbors, and to ensure no $\delta(v) \le d_i$ was updated. This proves Theorem~\ref{thm:numberofsubsteps}.

\subsection{Number of Steps}
\label{sec:step}
Let $\rho > 0$ be such that the condition $B(v,r(v))\ge \rho$ holds for all
$v\in V$.  We prove Theorem~\ref{thm:numberofrounds} by showing the following
lemma:


\begin{lemma}
\label{lem:round}
In any $t=1+\lceil\log_2 \rho L\rceil$ consecutive steps, except possibly in the
last step of the algorithm, we have $|S_{i+t}-S_i|\ge \rho$.
\end{lemma}

The lemma says that \AlgName{} will visit at least $\rho$ vertices in any
$t=1+\lceil\log \rho L\rceil$ consecutive steps, except possibly in the last
one. Hence, all vertices will be visited in no more than
$\lceil n/t\rceil=\lceil{ n/\rho}\rceil\left(1+\lceil \log_2 \rho
  L\rceil\right)$
steps.  

To prove this lemma, we show that in a step $j$, $i < j \leq i + t$, either
$\rho$ vertices have already been visited in the $j$-step
(i.e. $|S_{j+1}-S_{i}|\geq \rho$), or the difference in the step distance
doubles (i.e. $d_j-d_i\ge 2(d_{j-1}-d_i)$).  In the latter case, if that happens
for $t$ consecutive steps, then $d_t-d_i\geq 2^{t-1}>\rho L$, which has at least
$\rho$ nodes in this range.
We begin the proof by showing some basic facts about the algorithm: The lemma
below shows the property of the enclosed ball of the lead node $v_j$.

\begin{lemma}
\label{lem:neighbor}
If $r(v_j)<d(s,v_j)-d_i$, then $B(v_j,r(v_j))\subseteq (S_{j}-S_{i})$.
\end{lemma}

\begin{proof}
  Consider a vertex $v\in B(v_j,r(v_j))$.  First, $v$ cannot be visited after
  the $j$-th step, since $d(s,v)\le d(s,v_j)+d(v_j,v)\le \delta(v_j)+r(v_j)=d_j$.
Similarly, $v$ cannot be visited in the first $i$ steps since $d(s,v)\ge d(s,v_j)-d(v_j,v)>d_i+r(v_j)-d(v_j,v)\ge d_i$.
\end{proof}

The next property states that starting from the $i$-th step, either we reach
$\rho$ vertices, or the distance we explore in each step doubles.

\begin{lemma}
\label{lem:double}
If $|B(v_j,r(v_j))|\ge \rho$, then $\forall j>i$, either $|S_{j}-S_i| \ge \rho$, or $d_j-d_i\ge 2(d_{j-1}-d_i)\ge 2^{j-i-1}$.
\end{lemma}

\begin{proof}
Lemma~\ref{lem:neighbor} already shows that if $r(v_j)<d(s,v_j)-d_i$ then we have $|S_j-S_i|\ge |B(v_j,r(v_j))|\ge \rho$.  Otherwise, since $\delta(v_j)\ge d(s,v_j)$,
\begin{align*}
d_j-d_i&=\delta(v_j)+r(v_j)-d_i \\
&\ge d(s,v_j)+\left(d(s,v_j)-d_i\right)-d_i\\
&=2(d(s,v_j)-d_i)\ge 2(d_{j-1}-d_i),
\end{align*}
where the last step used the fact that $d(s,v_j)\ge d_{j-1}$
(Line~\ref{line:computeleadnode}).  Since the minimum edge weight is 1, we have
that $d_{i+1}- d_i\ge 1$ and further $v_j - v_i \ge 2^{j-i-1}$.
\end{proof}

With Lemma~\ref{lem:double}, we now prove Lemma~\ref{lem:round}, which is
equivalent to proving Theorem~\ref{thm:numberofrounds}.

\medskip {\sc Proof of Lemma~\ref{lem:round}.} {If $|B(v_j,r(v_j))|\ge \rho$ and $|S_{i+t}-S_i|<\rho$, then based on Lemma~\ref{lem:double}, $d_{i+t}-d_i\ge2^{t-1}\ge \rho L$.  Consider the node $u^*=\arg \min_{u\in V\backslash S_{i+t}} d(s,u)$. Let $\{u_1=s,u_2,\cdots,u_k=u^*\}$ be the ancestors to reach $u^*$ on the shortest-path tree.  From the definition of $u^*$ we have $\{u_1=s,u_2,\cdots,u_{k-1}\}\subseteq S_{i+t}$.  Meanwhile, since the longest edge in $G$ has edge weight $L$, for any $j$ that $k-\rho\le j\le k-1$, the shortest-path distance to $u_j$ satisfies that $d(s,u_j)\ge d(s,u^*)-(k-j)L\ge d(s,u^*)-\rho L>d_{i+t}-\rho L\ge d_i$, which means $u_j\notin S_i$.  Therefore, $\{u_{k-\rho},u_{k-\rho+1},\cdots,u_{k-1}\}\in S_{i+t}-S_i$, and $|S_{i+t}-S_i|\ge\rho$. \qed} 

\subsection{Work and Depth}
\label{sec:complexity}
\label{sec:workanddepth}

We now analyze the work and depth of \AlgName{}. The pseudocode of
an efficient implementation is shown in Algorithm~\ref{alg:efficientversion}.
Throughout the complexity analysis, we assume $|B(v,r(v))|\ge \rho$ and
$r(v)\le\rk{v}$ for all $v\in V$ (we show how to preprocess the graph to
guarantee this property in the next section).  

There are mainly two steps in Algorithm~\ref{alg:sssp} that are costly and
require efficient solutions: the calculation of the round distance $d_i$
(Line~\ref{line:computeleadnode}) and the discover of all unvisited vertices
with tentative distances less than $d_i$ (Line~\ref{line:innerloop}).

To efficiently support these two types of queries, we use two ordered sets $Q$ and $R$, implemented as balanced binary search trees (BSTs) to store the tentative distance ($\delta(u)$) for each unvisited vertex and the tentative distance plus the vertex radius ($\delta(u)+r(u)$).   With these two BSTs, the selection of round distance corresponds to finding the minimum in the tree $R$, and picking vertices within a certain distance is a split operation on the BST $Q$.  Both operations require $O(\log n)$ work.

Let $A_i$ be the \defn{active set} that contains the elements visited in the $i$-th step (all vertices $u$ in line~\ref{line:innerloop} in Algorithm~\ref{alg:sssp}).  This can be computed by splitting the BST $Q$ (Line~\ref{line:split} in Algorithm~\ref{alg:efficientversion}).   To maintain the priority queues sequentially, we relax all neighbors of vertices in $A_i$ for $k+2$ substeps.  Other than updating the tentative distance, the status of the neighbor vertex (referred to as $v$) may have three conditions:
\begin{enumerate}[label=(\arabic*),topsep=1pt,itemsep=0pt,parsep=0pt]
\item if the previous distance of $v$ is already no more than $d_i$, $v$ is
  already removed from $Q$ and $R$ in earlier this step, so we do nothing;
\item if the previous distance is larger than $d_i$ while the updated value is
  no more than $d_i$, we remove $v$ from $Q$ and $R$, and add $v$ to the active
  set $A_i$; and
\item if the updated distance is still larger than $d_i$, we decrease the keys
  corresponding to $v$ in $Q$ and $R$.
\end{enumerate}

\begin{lemma}[correctness]
  The efficient version with BSTs $Q$ and $R$
  (Algorithm~\ref{alg:efficientversion}) visits the same vertices in each step
  as Algorithm~\ref{alg:sssp} and computes the shortest-path distances
  correctly.
\end{lemma}

The proof of this lemma is straightforward.

The key step to parallelize Algorithm~\ref{alg:efficientversion} is to handle
the inner loop (Line~\ref{line:innerloop2}), which can be separated into two
parts: to update the tentative distances, and to update the BSTs.  We process
these two step one after the other.  To update the tentative distances, we just
use priority-write (\mf{WriteMin}) to relax the other endpoints for all edges
that have endpoints in $A_i$.  After all tentative distances are updated, each
edge checks whether the relaxation (priority-write) is successful or not.  If
successful, the edge will \emph{own} this vertex (called $u$), and create a BST
node that contains a pair of keys (lexicographical ordering): the current
tentative distance of $u$ as the first key, and the vertex label of $u$ as the
second key.  Then, we create a BST that contains all these nodes using standard
parallel packing and sorting process.  With this BST, we first apply
\mf{difference} operation to remove out-of-date keys in $Q$, then we \mf{split}
this BST into two parts by $d_i$, and \mf{union} each part separately with $A_i$
and $Q$.  It is easy to see that in this parallel version we apply
asymptotically the same number of operations to $Q$ compared to the sequential
version.  We can maintain $R$ in a similar way.

Now let us analyze the work and depth of the parallel version.  Each vertex is only in the active set in one round, so each of its neighbors is relaxed $k+2$ times, once in each substep.  In the worst case (all the relaxation succeeded), there are $O(m)$ updates for $Q$ and $R$, each can be done for no more than $O(k)$ times when $r(v)\le \rk{v}$.

Suppose there are $t$ steps and the number of updates in step $i$ is
$a_i$. Then, we have $\sum_{i=1}^t{a_i}=m$, and $t=O(\frac{n}{\rho}\log\rho L)$ as
$|B(v,r(v))|\le \rho$.  The work and depth for set \mf{difference} and \mf{union}
are $O(p\log q)$ and $O(\log q)$ for two sets with size $p$ and $q$,
$q \geq p$.  The overall work $W$ for all set operations is no more than
$\sum_{i=1}^{t}{a_i\cdot k\log n}=O\left(km\log n\right)$.
The depth
$D$ is no larger than
\[O\left(\sum_{i=1}^{t}{k\log n}\right)=O\left(k\frac{n}{\rho}\log \rho L\log n\right)\]

\hide{ Due to the concavity of the logarithm function, we have:
\begin{align*}
D&= O\left(k\log n\sum_i{\log a_i}\right)\\
&= O\left(k\log n \cdot {n\over \rho}\log \rho L\cdot{\log {m\over (n/\rho)\log \rho L}}\right)\\
&= O\left(k{n\over \rho}\log n \log{\rho m\over n} \log\rho L\right)
\end{align*}}
All other steps in Algorithm~\ref{alg:efficientversion} are dominated by the cost to maintain the BSTs. The total work and depth can be summarized in the following lemma.

\begin{lemma}
Assuming $|B(v,r(v))|\ge \rho$ and $r(v)\le\rk{v}$ for all $v\in V$, the \AlgName{} algorithm computes single-source shortest-path in $O\left(km\log n\right)$ work and\\ $O\left(k\frac{n}{\rho}\log n \log\rho L\right)$ depth.
\end{lemma}

Since $k$ can be set to be any positive integer in the preprocessing, from the theoretical point of view we can assume $k=1$ (larger $k$ is only for implementation purposes, and even then, $k$ is set to be no more than a small constant). This leads to $O\left(m\log n\right)$ work and $O\left(\tfrac{n}{\rho}\log n \log\rho L\right)$ depth. This indicates that we only have a $\log n$ factor of overhead compared to a standard Dijkstra's implementation. Along with the depth bound, this is the best work-depth tradeoff known for parallel \sssp{}.

\begin{algorithm}[t]
\caption{The Shortest-path Algorithm.}
\label{alg:efficientversion}
\KwIn{A graph $G=(V,E,w)$, vertex radii $r(\cdot)$ and a source node $s$.}
\KwOut{The graph distances $\delta(\cdot)$ from $s$.}
    \vspace{0.5em}
\DontPrintSemicolon
$\delta(\cdot)\leftarrow +\infty$, $\delta(s)\leftarrow 0$\\
$i\leftarrow 1$\\
$Q=\{w(s,u)\,|\,u\in N(s)\}$\\
$R=\{w(s,u)+r(u)\,|\,u\in N(s)\}$\\
\While{$|Q|>0$\label{line:loop}} {
$d_{i}\leftarrow R.\mf{extract-min}()$\\
$\{A_i,Q\}=Q.\mf{split}(d_i)$\label{line:split}\\
\lForEach {$u\in A_i$} {$R.\mf{remove}(u)$}
\Repeat{no $\delta(v)$ that $v\in A_i$ is updated\label{line:substeploop}} {
    \ForEach{$u\in A_i, v\in N(u)$\label{line:innerloop2}} {
    \If {$\delta(v)>\delta(u)+w(u,v)$} {
    \If {$\delta(v)>d_i$ \emph{and} $\delta(u)+w(u,v)\le d_i$} {
    $R.\mf{remove}(v)$\\
    $Q.\mf{remove}(v)$\\
    $A_i.\mf{insert}(v)$
    }
    $\delta(v)=\delta(u)+w(u,v)$\\
    \If {$\delta(v)>d_i$} {
    $Q.\mf{decrease-key}(v,\delta(v))$\\
    $R.\mf{decrease-key}(v,\delta(v)+r(v))$
    }
    }
}
}
$i=i+1$\\
}
\Return {$\delta(\cdot)$}\\[.25em]

\hide{
  \SetAlgoLined
  \SetKwFunction{algo}{algo}\SetKwFunction{funcName}{update}
  \SetKwProg{myfunc}{function}{}{}
  \myfunc{\mf{update}\emph{(vertex $u$ and $v$, distance $w$)}}{
\If {$\delta(v)>\delta(u)+w$} {
\If {$\delta(v)>d_i$ \emph{and} $\delta(u)+w\le d_i$} {
$R.\mf{remove}(v)$\\
$Q.\mf{remove}(v)$\\
$A_i.\mf{insert}(v)$
}
$\delta(v)=\delta(u)+w$\\
\If {$\delta(v)>d_i$} {
$Q.\mf{decrease-key}(v,\delta(v))$\\
$R.\mf{decrease-key}(v,\delta(v)+r(v))$
}
}
  }
  }
\end{algorithm}

\subsection{Unweighted Cases}
\label{sec:unweighted}
The implementation we just discussed works on unweighted cases as well. However,
the performance can be further improved in this case. A special property of the
unweighted case is that all vertices in the frontier have the same tentative
distances.  This means that we do not need search trees or priority queues to
maintain the ordering.  Hence, a similar approach to parallel BFS can be
directly used here, and each round takes $O(n')$ work and $O(\log^* n')$ depth
on the CRCW PRAM model \footnote{The bound changes with different models.} where
$n'$ is the number of vertices and their associated edges in each round.  The
additional step to compute the round distance uses one priority-write.  Because
of concavity, the worst case appears when $n'$ is the same in every round.

\begin{lemma}\label{lem:efficiency}
The \AlgName{} algorithm computes single-source shortest-path on unweighted graph in $O(m+n)$ work and $O\left((n/\rho)\log\rho\log^*\rho\right)$ depth.
\end{lemma} 


\section{Shortcuts and Preprocessing}
\label{sec:ball}

This section describes how to prepare the input graph so that \AlgName{} can run
efficiently afterward.  The cost of \AlgName{}, as we proved earlier, is a
function of $k$ and $\rho$, where the input graph to the algorithm and the
$r(\cdot)$ function have to satisfy $r(v) \leq \rk{v}$ and
$|B(v,r(v))| \geq \rho$ for all $v \in V$.

Not all graphs as given meet the conditions above with settings of $k$ and
$\rho$ that yield plenty of parallelism.  Moreover, for a given $k$, directly
computing $\bar{r}_k(v)$ is expensive; it may require as much as $O(nm)$
work.

Our aim is therefore to add a small number of extra edges (shortcuts) to satisfy
the conditions for $k$ and $\rho$ that the user desires.  We will also generate
an appropriate vertex radius $r(v)$ for every $v\in V$.  We begin with the definitions of $\rho$-nearest distance and
\ourstructure{k}{\rho}:



\begin{definition}[$\rho$-nearest distance]
  For $v \in V$, the $\rho$-nearest distance of $v$, denoted by $r_\rho(v)$, is
  the distance from $v$ to the $\rho$-th closest vertex to $v$.
\end{definition}

\begin{definition}[\ourstructure{k}{\rho} and \ourgraph{k}{\rho}]
  We say that a vertex $v\in V$ has a \ourstructure{k}{\rho} if
  $r_\rho(v)\le \rk{v}$.  A graph is a \ourgraph{k}{\rho} if each vertex in the
  graph forms a \ourstructure{k}{\rho}.
\end{definition}

The following lemma is an immediate consequence of the definitions:

\begin{lemma}\label{lem:link}
  If $r(v) = r_\rho(v)$ for all $v \in V$ and the input graph to the \AlgName
  algorithm is a \ourgraph{k}{\rho}, then $r(v) \leq \rk{v}$
  and $|B(v,r(v))| \geq \rho$.
\end{lemma}

This means that the numbers of steps and substeps of the \AlgName{} algorithm
are bounded as stated in Theorems~\ref{thm:numberofsubsteps} and
\ref{thm:numberofrounds}.

It is easy to convert any graph into a \ourgraph{k}{\rho} by adding up to
$n\rho$ edges (put shortcut edges from every vertex to its $\rho$-nearest
vertices).  For $k = 1$, this strategy adds the fewest number of edges.
For $k > 1$, however, there is room for a better scheme.  Below, we
discuss efficient preprocessing strategies and their cost.






\subsection{Heuristics for \ourstructure{1}{\rho}s}

The simplest case of a \ourstructure{k}{\rho} is the \ourstructure{1}{\rho}.  In
this case, all the $\rho$-closest vertices from a vertex $u$ are directly added
to $u$'s neighbor list with edge weight $d(u,\cdot)$.

To generate the \ourstructure{1}{\rho}s for all the vertices, we can, in
parallel, start $n$ Dijkstra's execution (or BFS for unweighted graphs) and
compute the $\rho$-closest vertices in each run.

\begin{lemma}
\label{lem:preprocessing}
Given a graph $G$, generating \ourstructure{1}{\rho}s for all vertices takes
$O(m\log n+n\rho^2)$ work and $O(\rho^2)$ depth, or $O(m\log n+n\rho^2\log \rho)$
work and $O(\rho \log \rho)$ depth.
\end{lemma}

\begin{proof}
We initially sort all edges from each vertex by their weights, requiring $O(m\log n)$ work and $O(\log n)$ depth.  This step can be saved if the edges are presorted or the graph is unweighted.  Then, we run, in parallel, standard Dijkstra's algorithm~\cite{Dijk} from each vertex for $\rho$ rounds, and use Fibonacci Heap~\cite{Fib} as the priority queue.  However, for each vertex, we only consider the lightest $\rho$ edges since only these edges are sufficient to reach the $\rho$-closest vertices from each source node.  The search from each vertex therefore explores no more than $\rho^2$ edges ($\rho^2$ {\sc Decrease-key}s in the Fibonacci Heap) and visits $\rho$ vertices ($\rho$ {\sc Delete-min}s in the Fibonacci Heap), leading to $O(\rho^2)$ work for each node. In total, this is  $O(\rho^2)$ depth and $O(n\rho^2)$ work for the Dijkstra's component.

We can then add $\rho$ shortcut edges from the source to all these vertices.
When the algorithm stops in the $\rho$-th round, we acquire $r_\rho(v)$, the distance from the source to $\rho$-th nearest neighbor directly.  Clearly $r_\rho(v)\le \hat{r}_1(v)$ at this time.

The priority queue can also be implemented with a balanced BST, and similar to the discussion in Section~\ref{sec:workanddepth}, such process takes $O(m\log n+n\rho^2\log \rho)$ work and $O(\rho \log \rho)$ depth.
\end{proof}

For unweighted case, a parallel BFS from the source vertex to reach $\rho$-nearest neighbors takes $O(\rho^2)$ work and $O(\rho\log^*\rho)$ depth.

Combining Lemmas~\ref{lem:preprocessing},~\ref{lem:link},
and~\ref{lem:efficiency} proves the main theorem of this paper
(Theorem~\ref{thm:mainthm}).  After preprocessing, the graph now has
$O(m+n\rho)$ edges.

\begin{figure}[t]
\begin{center}
  \includegraphics[width=0.6\columnwidth]{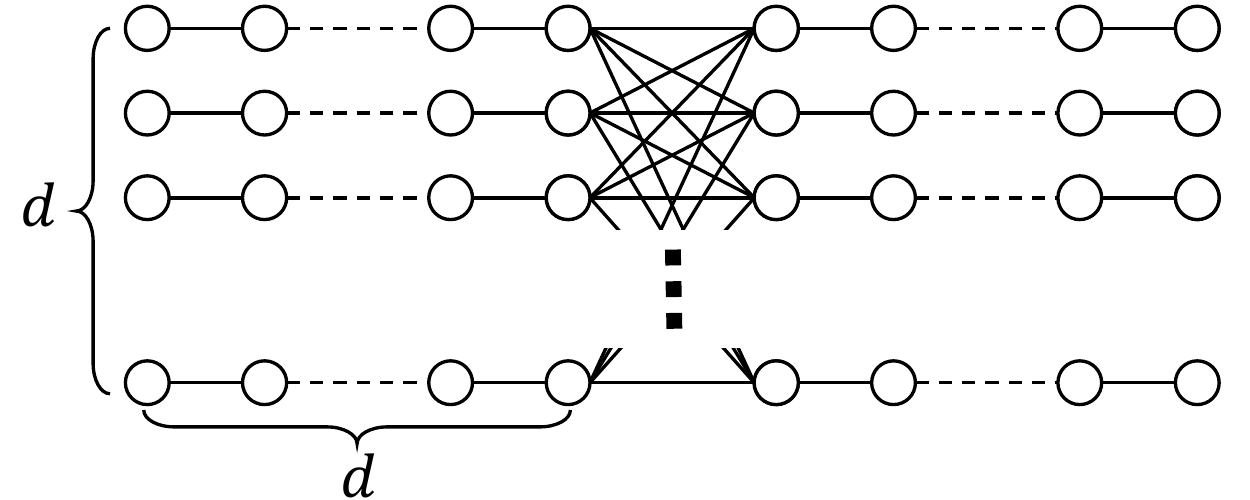}
  \caption{Example of a sparse graph that requires BFS to look at $O(d^2)$ edges
    from any vertex to reach $3d$ vertices.}\label{fig:construction}
\end{center}
\end{figure}

It is worth mentioning that even given an unweighted sparse graph, we may still
need to look at $O(\rho^2)$ edges to reach $\rho$ vertices.  We give an
example of a carefully-constructed graph with that behavior in
Figure~\ref{fig:construction}. If $d=\lfloor\rho /3\rfloor-1$, then the BFS
search from any vertex has to visit $O(d^2)$ edges to reach $\rho>3d$ vertices.
However, in Section~\ref{sec:exp} we will show via experiments that this case is
not typical in real-world graphs.  Furthermore, if the input graph has constant
degree for each node, the work for this step is $O(n\rho)$.

\subsection{Heuristics for \ourstructure{k}{\rho}s}
\label{sec:heuristics}

The construction of a \ourgraph{1}{\rho} requires up to $n\rho$ extra edges,
  potentially too many to be useful for many graphs in practice.  Using a small $k > 1$
  creates an opportunity to add fewer shortcut edges, by taking better advantange of the
  original edges.




To construct a \ourstructure{k}{\rho}, we still need to run
the algorithm mentioned in Lemma~\ref{lem:preprocessing}, after
which a shortest-path tree with $\rho$ vertices can be built.
Consider the shortest-path tree $T_s$ starting from a source $s \in V$ that
reaches out to the $\rho$ nearest vertices.
Our goal is to add as few edges as possible to ensure that every $v \in T_s$ can
be reached within $k$ hops using the combination of the original edges and the
added edges $E'$.

The following claim, which is easy to prove, shows that from the perspective of
a single source vertex, the most beneficial shortcut edge goes directly from the
source itself.
\begin{claim}
  An edge $(u, v)$, $u\neq s$, can be exchanged for $(s, v)$ without increasing
  the overall hop count from $s$.
\end{claim}
We propose two heuristics to construct \ourstructure{k}{\rho}s.

\subsubsection{Shortcut using Greedy}
To directly eliminate vertices that are more than $k$ hops away in $T_s$, we add
edges from the root to all $(ki+1)$-hop neighbors for $i\in \mathbb{Z}^+$. This
heuristic is simple and easy to program, but may result in adding too many
edges. Consider a shortest-path tree that first forms a chain of length $k$ from
the source, then the rest $\rho-k-2$ vertices all lie in the $(k+1)$-th
level. The greedy heuristic will shortcut to all these $\rho-k-2$ vertices,
while we can only add one edge from the source to the node in the $k$-th layer
to include all vertices in $k$ hops.

\subsubsection{Shortcut using Dynamic Programming (DP)}
Consider, among all shortest-path trees from $s$, one where for every
$v \in T_s$, the path $s \to v$ on $T$ has the smallest hop count possible. The
optimal number of edges needed for a certain node can be computed using the
following dynamic program. Let $F(u, t)$ be the number of edges into the subtree
of $T_s$ rooted at $u$ so that every node in the subtree is at most $k$ hops
away from $s$ provided that $\mbox{parent}(u)$ is at $t$ hops away from $s$.
Then, for all $u \neq s$, $F(u, t)$ is given by
\begin{align*}
   \begin{cases}
    1 + \sum\limits_{w \in N^+(u)} F(w, 1) & \text{ if } t = k\\
    \min\left(1 + \sum\limits_{w \in N^+(u)} F(w, 1),
      \sum\limits_{w \in N^+(u)} F(w, t+1)\right)  & \text{ if } t < k\\
  \end{cases}
\end{align*}

The number of edges needed in the end is $\sum_{u \in N^+(s)} F(u ,0)$.  The
actual edges to add can be traced from the recurrence.  We note that this
dynamic program can be solved in $O(k\rho)$ (as $|T_s| = \rho$) by resolving the
recurrence bottom up (using either DFS or BFS).



Note that although the dynamic programming solution gives an optimal solution for each
shortest-path tree individually, the overall solution is not necessary the
global optimal.
Adding globally smallest number of edges to construct \ourstructure{k}{\rho}
seems much more difficult.  Yet, we show empirically in Section~\ref{sec:exp}
that the dynamic programming (DP) heuristic performs very well on a wide variety
of graphs---and even the greedy heuristic does well on well-structured graphs.




\section{Experimental Analysis}
\label{sec:exp}

This section empirically investigates how the settings of $k$ and $\rho$, and
the choice of shortcutting heuristics affect the number of shortcut edges added,
as well as the number of steps that \AlgName{} requires.  These
quantities provide an indication of how well a well-engineered implementation
would perform in practice.



\subsection{Experiment Setup}
We use graphs of various types from the SNAP datasets~\cite{leskovec2014snap},
including the road networks in Pennsylvania and Texas (real-world planer graphs)
and the web graphs of the University of Notre Dame and Stanford University
(real-world networks).  In the case of web graphs, each edge represents a
hyperlink between two web pages.  We also use synthetic graphs of 2D and 3D
grids (structured and unstructured grids).

In practical applications, such graphs are used both in the weighted and
unweighted settings.  In the weighted setting, for instance, the distances in
road networks and grids can represent real distances, while the edge weights in
network graphs (e.g., social networks) also have real-world meanings, such as
the time required to pass a message between two users, or (the logarithm of) the
probability to pass a message.  In our experiments, if a graph does not come
equipped with weights, we assign to every edge a random integer between $1$ and
$10,000$.

We construct \ourstructure{k}{\rho}s using the heuristics described in
Section~\ref{sec:heuristics} except the implementation has the following
modifications: instead of breaking ties arbitrarily and taking exactly $\rho$
neighbors, we continue until all vertices with distance $r_\rho(\cdot)$ are
visited.  This has the same implementation complexity as the theoretical
description but is more deterministic.  Using this, our results are a
pessimistic estimate of the original heuristics as more than $\rho$ edges may be
found for some sources.  However, in all our experiments, we found the
difference to be negligible in most instances.

To improve our confidence in the results, two of the authors have independently
implemented and conducted the experiments, and arrived at the same results, even
without introducing a tie-breaker.

\subsection{The Number of Shortcut Edges}

Section~\ref{sec:heuristics} described heuristics that put in a small number of
edges to make the input graph a \ourgraph{k}{\rho}.  \emph{How many extra edges
  are generated by each heuristic?}  To answer this question, we use 3
representative graphs in our analysis: (1) road networks in Pennsylvania, (2) a
webgraph of Stanford University, and (3) a synthetic $1000$-by-$1000$ 2D grid.
All these graphs have about $1$ million vertices, and between $2$ and $3$
million edges.  We show experimental results for unweighted graphs since the
performance of the heuristics is independent of edge weights.

\begin{figure}[!t!h]
\begin{center}
    \begin{minipage}[t]{0.45\textwidth}
        \includegraphics[width=\textwidth]{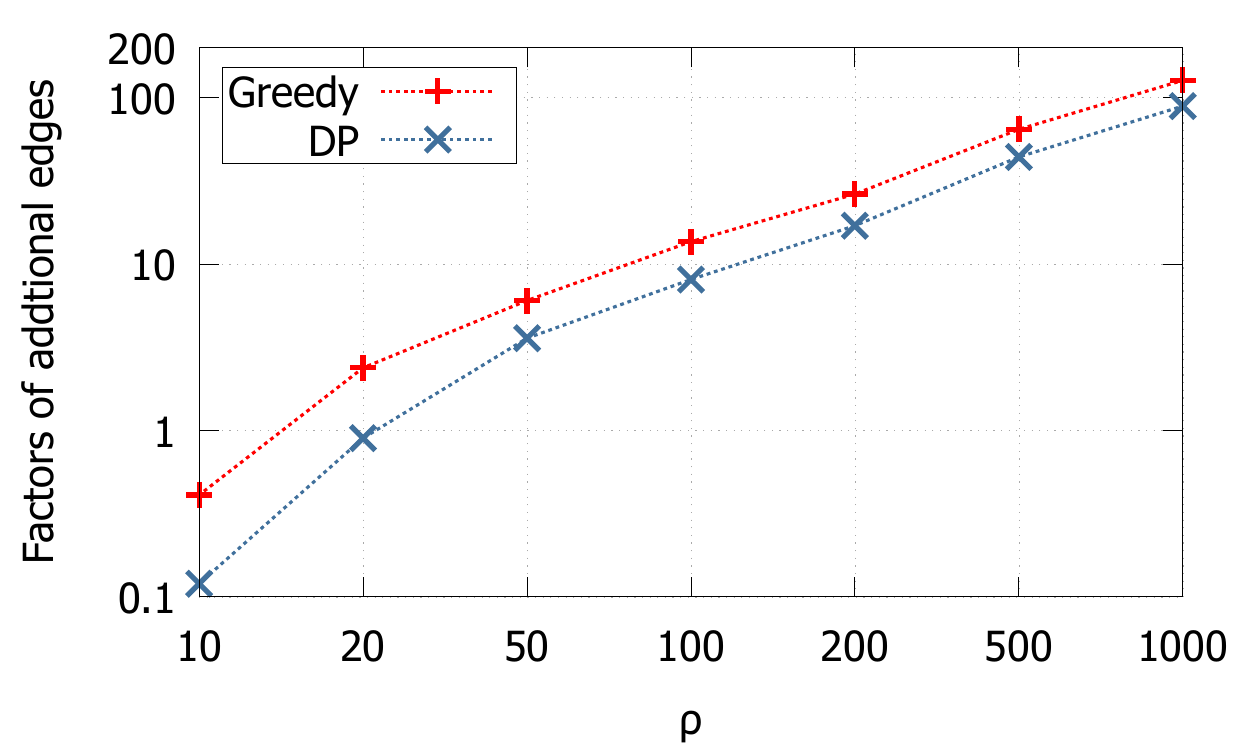}\\ \centering{\small \bf (a) Road map of Pennsylvania}
    \end{minipage}
    \begin{minipage}[t]{0.45\textwidth}
        \includegraphics[width=\textwidth]{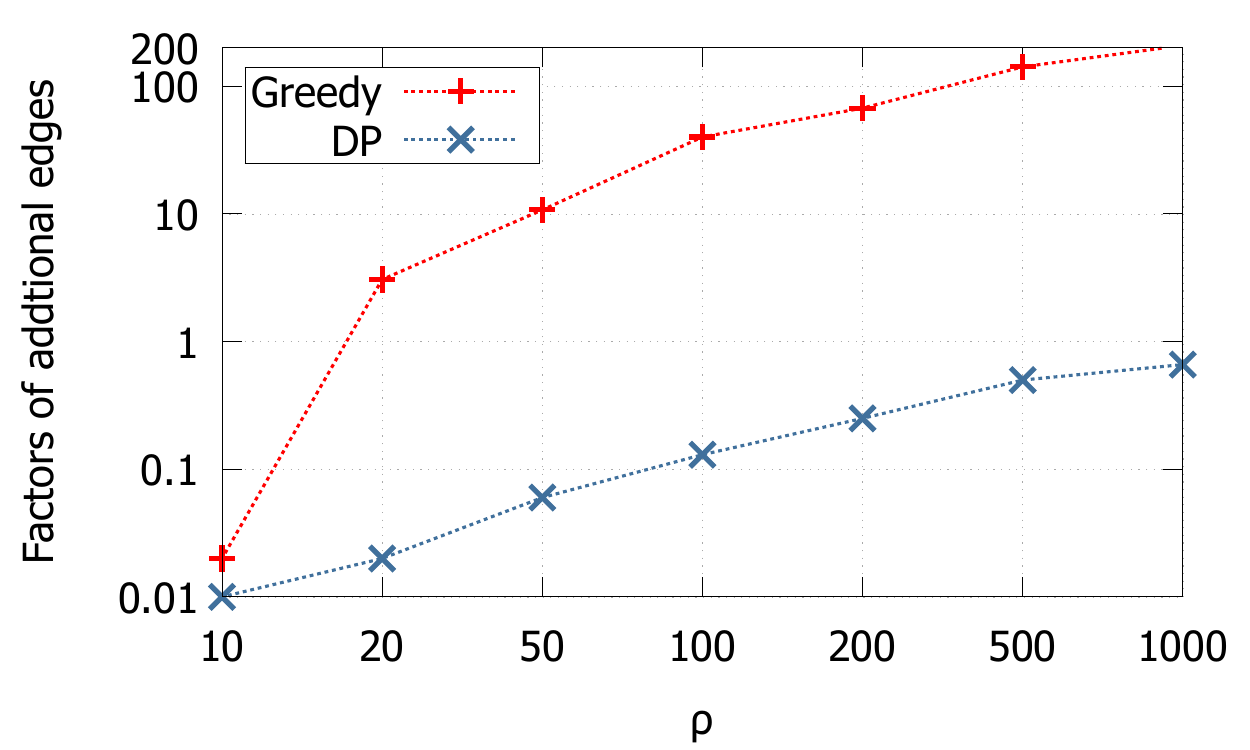}\\ \centering{\small \bf (b) Webgraph of Stanford}
    \end{minipage}
    \begin{minipage}[t]{0.45\textwidth}
        \includegraphics[width=\textwidth]{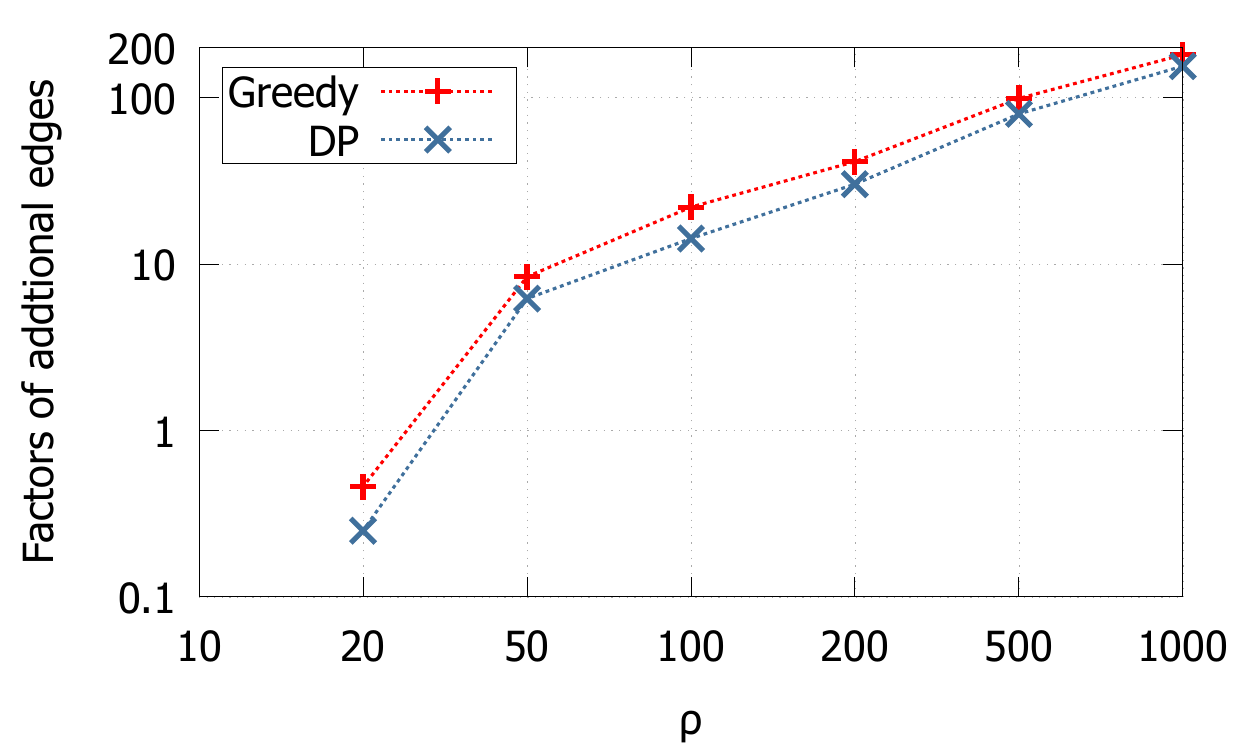}\\ \centering{\small \bf (c) 2D-grid}
    \end{minipage}
    \caption{The fraction of additional edges added to generate
      \ourgraph{k=3}{\rho} using greedy vs. dynamic programming (DP) for different
      types of graphs.}\label{fig:addingedges}
\end{center}
\end{figure}

Figure~\ref{fig:addingedges} shows the number of added edges in terms of the
fraction of the original edges for $k = 3$ as the value of $\rho$ is varied
between $10$ and $1,000$.  More detailed results are given in
Tables~\ref{tab:addingedgegreedy} and \ref{tab:addingedgedp} in the appendix.
Evidently, both heuristics achieve similar results on the road map and 2D
grid. This is because road maps and grids are relatively regular, in fact
almost planar.  Thus, even the na\"ive shortcuts to the $(ki+1)$-hop performs
well. As $k$ becomes larger, the gap between the greedy and the DP heuristic
increases.  When $\rho$ reaches $1000$, both DP and greedy add more than $100$x
edges compared to the original graph. This is because in road maps and grids the
degree of a vertex is usually a small constant, which makes the shortest-path tree very
deep.

On webgraphs, which is less regular, DP only adds $4$x the number of original
edges even when $\rho$ is as large as $1000$ while greedy still adds $100$x the
original edges. This is because webgraphs not only are far more irregular but
also have a very skewed degree distribution.  As a known scale-free network
\cite{barabasi1999emergence}, it has some ``super stars'' (or more precisely,
the ``hubs'') in the network.  In this case, the bad example in
Section~\ref{sec:heuristics} occurs frequently when these hubs are not at the
exact $(ki+1)$-th layer in the shortest-path tree, while the DP heuristic can
discover the hubs accurately.  This also explains the phenomena that only a few
edges are added on webgraphs by the DP heuristic even when $\rho$ is large,
since the hubs already significantly reduce the depth of the shortest path tree,
and it only takes a few edges to shortcut to the hubs.  This property holds for
many kinds of real-world graphs such as social networks, airline networks,
protein networks, and so on.  In such graphs, a relatively optimal heuristic is
necessary to construct the enclosed balls; a na\"ive method often leads to bad
performance.  As can be seen, the DP solution achieves satisfactory performance
on webgraphs, where it only adds about 10\% more edges with $k=3$ and
$\rho=100$.

\hide{
\kanat{Do we really want to mention the following? It seems very subtle and may
  muddle the point.}  Note that to efficiently implement it is important to
break the ties and create \emph{no more than} $\rho$ shortcuts. In that case the
number of added edges is at most $n\rho$.  From Table \ref{tab:addingedgedp} we
can see that even ties are not broken, the number of edges created by DP is
still much less than $n\rho$.
}

\subsection{The Number of Steps}
\label{sec:expreduce}

\emph{How many steps does \AlgName{} take for each setting of $\rho$?}  We ran
our \AlgName{} on both weighted and unweighted graphs and counted the number of
steps as we change $\rho$.  Notice that the number of steps is independent of
$k$ and is only affected by $\rho$. (The value of $k$ only affects the number of
substeps within a step.)  We discuss the performance of our algorithm on all six
graphs.  Since the cost of \sssp potentially varies with the source and we
cannot afford to try it from all possible sources, we take 1000 random sample
sources for each graph.  We use the same $1000$ sources for all our experiments
for both the weighted and unweighted cases.  We report the arithmetic means over
all sample sources.


\medskip

\begin{figure*}[t]
\begin{center}
    \begin{minipage}[t]{0.45\textwidth}
\begin{center}
    \begin{minipage}[t]{\textwidth}
        \includegraphics[width=\textwidth]{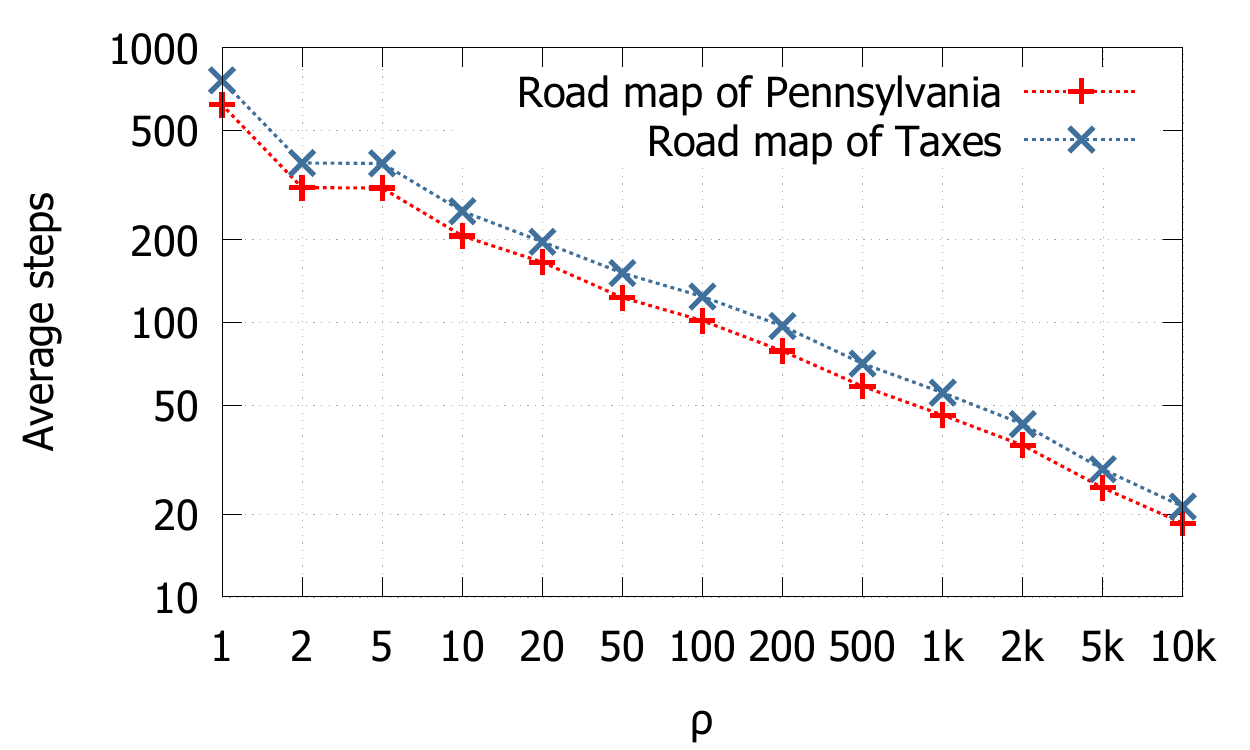}\\ \centering{\small \bf (a) Road maps}
    \end{minipage}
    \begin{minipage}[t]{\textwidth}
        \includegraphics[width=\textwidth]{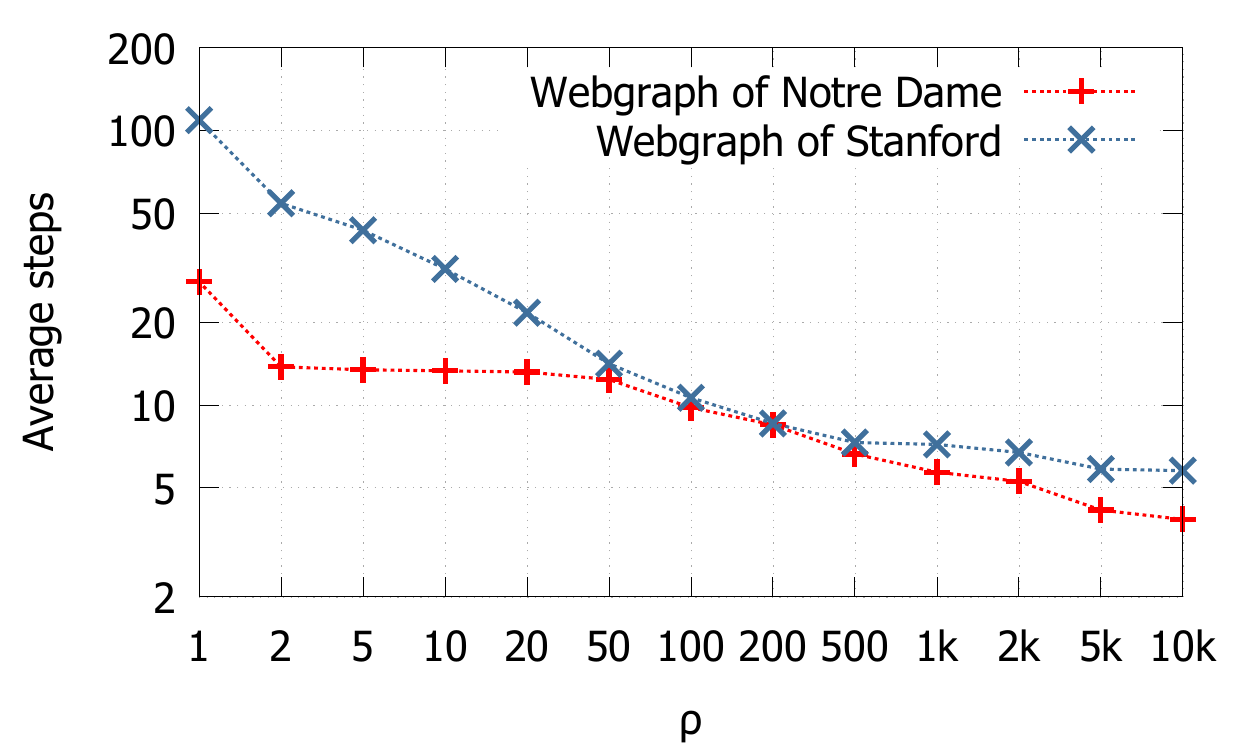}\\ \centering{\small \bf (b) Webgraphs}
    \end{minipage}
    \begin{minipage}[t]{\textwidth}
        \includegraphics[width=\textwidth]{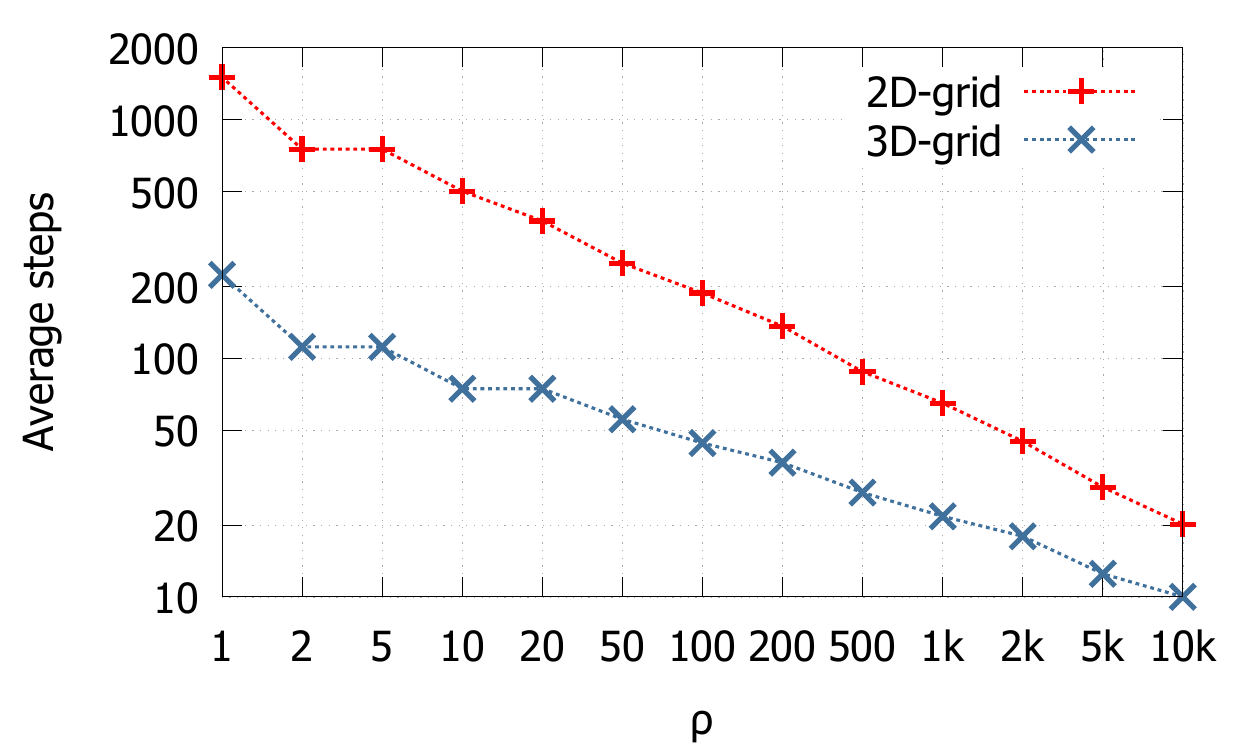}\\ \centering{\small \bf (c) Grids}
    \end{minipage}
  \caption{Unweighted graphs---the number of \AlgName{} steps as $\rho$ is varied.}\label{fig:numerofrounds}
\end{center}
    \end{minipage}  ~~~~~
    \begin{minipage}[t]{0.45\textwidth}
\begin{center}
    \begin{minipage}[t]{\textwidth}
        \includegraphics[width=\textwidth]{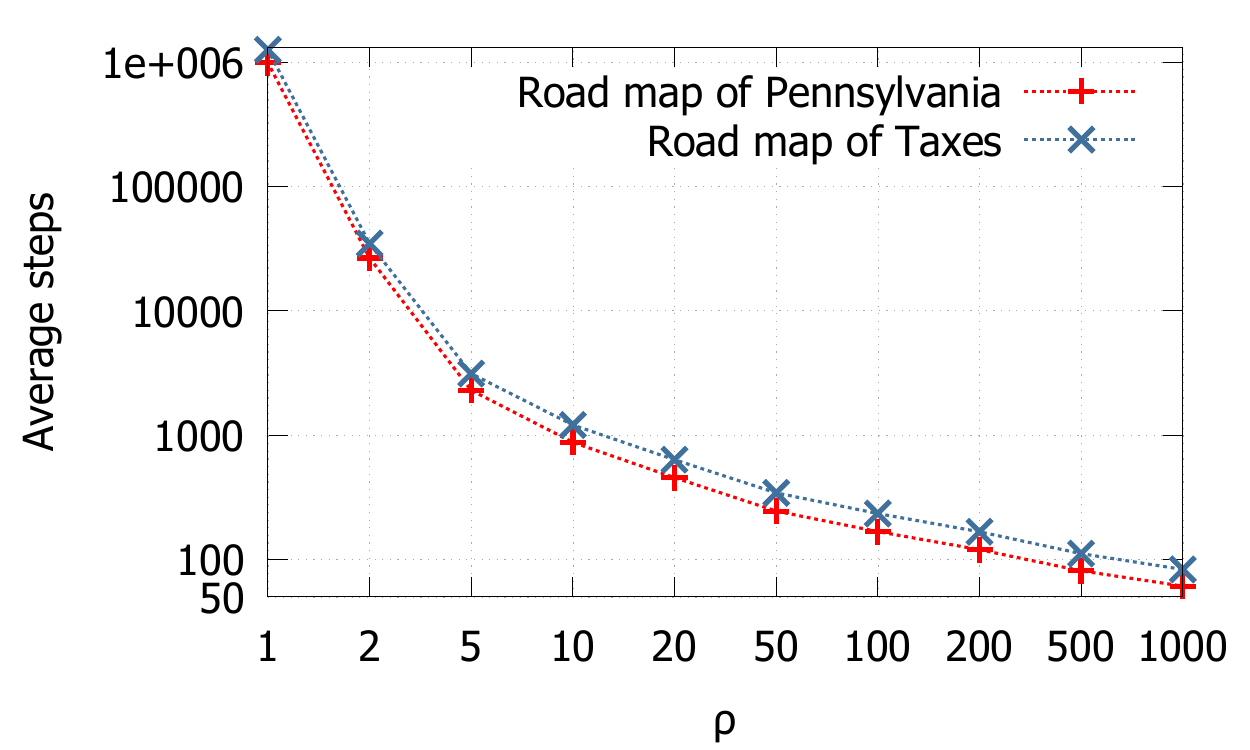}\\ \centering{\small \bf (a) Road maps}
    \end{minipage}
    \begin{minipage}[t]{\textwidth}
        \includegraphics[width=\textwidth]{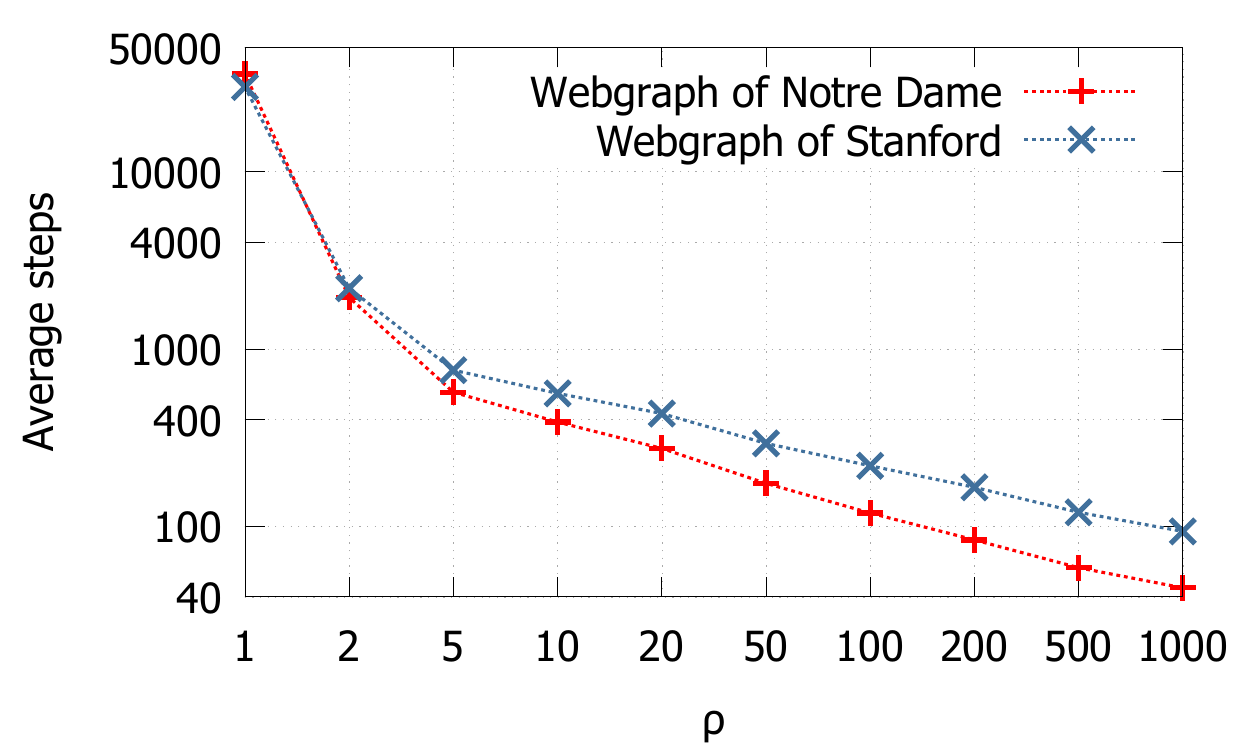}\\ \centering{\small \bf (b) Webgraphs}
    \end{minipage}
    \begin{minipage}[t]{\textwidth}
        \includegraphics[width=\textwidth]{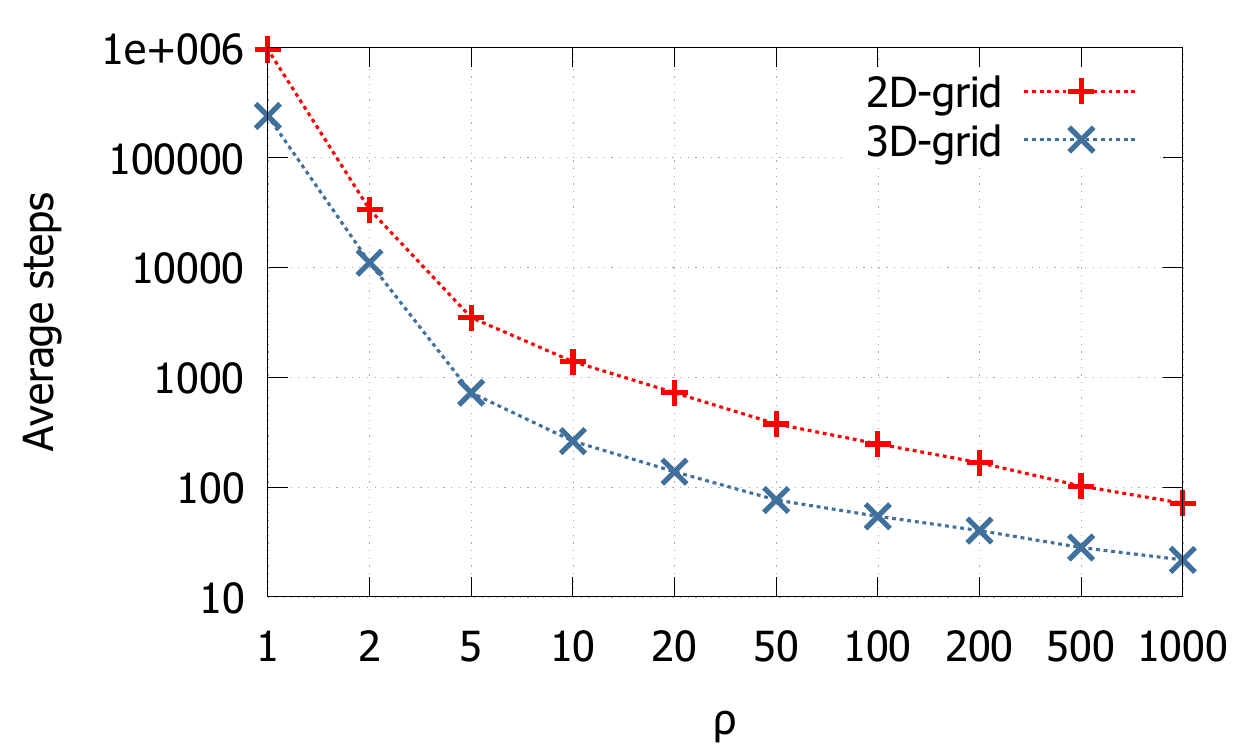}\\ \centering{\small \bf (c) Grids}
    \end{minipage}
    \caption{Weighted graphs---the number of \AlgName{} steps as $\rho$ is
      varied.}\label{fig:numerofroundsw}
\end{center}
    \end{minipage}
\end{center}
  \vspace{-2.5em}
\end{figure*}

\noindent
\textbf{Unweighted Graphs (BFS):} Figure~\ref{fig:numerofrounds} shows, for the
unweighted case, the average number of steps taken by \AlgName{} as $\rho$ is
varied.  More results appear in Tables~\ref{tab:stepunweighted} and
\ref{tab:stepreducedunweighted} in the appendix, which compare the number of
steps taken by \AlgName with that of a conventional BFS implementation.

Several things are clear: on a log-log scale\footnote{The vertical axis is drawn
  on a log scale and the horizontal axis closely approximates a log scale.}, the
trends are downward linear as $\rho$ increases except for the Notre Dame
webgraph (not completely regular but shows a similar trend).  This suggests that
the average number of steps is inversely proportional to $\rho$, which is
consistent with our theoretical analysis.

The webgraph has a relatively smoother slope. The reason is that once the
``super stars'' are included in the enclosed balls, which usually only need a few
hops, then most of the vertices will be visited in a few steps ($20$--$100$
steps vs. $200$--$1,500$ steps on the other graphs when $\rho=1$).  However, we
will later see that in the weighted case, the performances on these graphs are
even better than other graphs.  For the road maps and grids, the number of steps
reduces steadily. Furthermore, the number of steps shown in the experiments is
much less than the theoretical upper bound ($\frac{n}{\rho}\log \rho$) because
most real-world graphs tend to have a hop radius that is smaller than $O(n)$.



On all the graphs studied, $\rho$ can be as small as $10$ to reduce the number
of steps by $3$x.  When $\rho=100$, the reduction factor is about $10$x.  As the
results show, we can achieve a reduction factor of more than $20$x when thousands of
vertices are in the balls.
The experiment results verified our theoretical analysis of the \AlgName{} algorithm.

\emph{What should an unweighted graph look like so that \AlgName{} reduces the
  steps much when adding no more than $m$ edges?}  At first thought, the answer
might be a grid or a road map with a large diameter, so that there is more space
to be reduced.  However, our experiments give the opposite answer: \AlgName{},
in fact, performs better on webgraphs with smaller diameters.  On webgraphs,
\AlgName{} can reduce the number of steps by $15$x by adding no more than $m$
edges (choosing $\rho=1000$ and $k=3$), while on road maps and grids, a
5x reduction in steps requires $4m$ to $6m$ edges. Even though the number
of rounds reduces more steadily and quickly on road maps and grids with
larger $\rho$, the number of added edges in turn increases more rapidly
(100x times more edges added to achieve 20x reduction on the number of steps).
On scale-free networks, however, \AlgName{} improves standard BFS by more than
10x in time without adding much extra edges, and an efficient implementation of
\AlgName{} on these networks might be worthwhile in the future.

\medskip

\noindent\textbf{Weighted Graphs:}
Figure~\ref{fig:numerofroundsw} shows, for the weighted case, the average number
of steps taken by \AlgName{} as $\rho$ is varied.  More results appear in
Tables~\ref{tab:stepweighted} and \ref{tab:stepreducedweighted} in the appendix,
which compare the number of steps taken by \AlgName to when $\rho = 1$.  Notice
that when $\rho = 1$, \AlgName{} becomes essentially Dijkstra's except vertices with the same distance are extracted together.


Similar to the unweighted case, the trends in Figure \ref{fig:numerofroundsw}
are also nearly-linear, indicating an inverse-proportion relationship between
the number of steps and $\rho$, which is consistent with our theory.  In the
weighted case, however, we can visit much fewer vertices in each step than those
in the unweighted case because most vertices have different distances to the
source.  Thus, Dijkstra's algorithm (or when $\rho = 1$) requires almost $n$
steps to finish.  As a result, considering the inverse-proportion relationship,
even a small $\rho$ can reduce greatly the number
of steps.  Indeed, the number of steps taken by \AlgName is far fewer than the
predicted theoretical upper bound ($O(\frac{n}{\rho}\log \rho L)$ with $L=10^4$
here).  We also notice that the number of steps decreases much faster when
$\rho$ is small compared to larger values of $\rho$.


The reduction in the number of steps is significant. Even $\rho=10$ leads to
about $1000$x fewer steps on road maps and grids, and about $50$--$100$x on the
webgraphs. When $\rho=100$, we only need a few hundreds of steps on all graphs, which
can already provide considerable parallelism.  The number of steps further
decreases to $20$--$80$ when $\rho=1000$.  This provides some evidence that the
algorithm has the potential to deliver substantial speedups when many more cores
on a shared-memory machine are made available.

Another trend is that the reduction factors on the webgraphs are always less
than the other two types of graphs.  The reason is that the webgraphs are
scale-free, and even when we do not use enclosed balls to traverse the graph,
not many steps are required. The ``hubs'' substantially bring down the diameter
of the graph.  For more uniformly distributed graphs, such as the road maps and
grids, increasing $\rho$ to a big value reduces the number of steps required to
perform \sssp{} more rapidly.

\subsection{How to choose the parameters?}

\emph{What is the best combination of $k$ and $\rho$?}  The choice of $k$ and
$\rho$ offers a tradeoff between the number of added edges (hence additional
space and work) and the parallelism of the algorithm.  In general, we do not
want to increase the number of edges substantially; the total number of edges
should be around $O(m)$.  On every graph tested, $k = 3$ or $4$ works reasonably
well whereas $\rho$ in the range $50$--$100$ for weighted graph
yields the best bang for the buck.  A larger $\rho$ can reduce
the number of steps in \AlgName{} but in turn, increases the preprocessing time
and the number of edges required to be added. A larger $k$ will reduce the
number of added edges but at the same time, increases the number of visited
edges, as well as the overall depth\footnote{As mentioned at the beginning of
  Section~\ref{sec:expreduce}, a larger $k$ will not increase the number of
  steps, but the number of the overall substeps, which is the inner loop of
  \AlgName{}, hence an increase in the overall depth.}.

On unweighted graphs, \AlgName{} performs better on webgraphs, but less efficient on grids and road maps.
Finally, since
preprocessing is only run once, if \sssp will be run from multiple sources, we
suggest increasing $\rho$ and decreasing $k$: the cost for preprocessing is
amortized over more sources.


\section{Prior and Related Work}
\label{sec:related-work}

\begin{table*}[t]
  \centering\footnotesize
  \def\arraystretch{1.5}
  \begin{tabular}{p{0.5in}p{1.25in}p{1.6in}l p{1.2in}}
    \toprule
    \textbf{Setting} & \textbf{Algorithm} & \textbf{Work} & \textbf{Depth} & \textbf{Parameters}\\
    \midrule
    \textsf{Unweighted (BFS)} & Standard BFS & $O(m+n)$ & $O(n)$ \\
                     &Ullman and Yannakakis~\cite{ullman1991high} & $\otilde(m\sqrt{n} + nm/t + n^3/t^4)$ & $\otilde(t)$ & $t \leq \sqrt{n}$\\
                     &Spencer~\cite{spencer1997time} & $O(m\log \rho +n\rho^2\log^2 \rho)$ & $O(\tfrac{n}\rho \log^2 \rho)$ & $\sqrt{m/n} \leq \rho \leq n$\\
                     & \emph{\textsf{This work}} & $O(m+n\rho)$ & $O(\tfrac{n}\rho \log\rho \log^*\rho)$ & \hide{$\rho\le \sqrt{n}$}\\
                     & & preproc: $O(n\rho^2)$ & $O(\rho\log^* \rho)$\\
    \midrule
    \textsf{Weighted} \sssp & Parallel Dijkstra's~\cite{PaigeK:icpp85} & $O(m + n\log n)$ & $O(n\log n)$ \\
                     & Parallel Dijkstra's~\cite{brodal1998parallel} & $O(m\log n + n)$ & $O(n)$ \\
                     & Klein and Subramanian~\cite{klein1997randomized} &  $O(m\sqrt{n}\log K \log{n})$ & $O(\sqrt{n} \log K \log{n})$
                                                                           &  $K=\,$max dist.~from $s$\\
                     & Spencer~\cite{spencer1997time} & $O((n\rho^2\log \rho+m)\log (n\rho{}L))$
                                                          & $O(\tfrac{n}\rho \log n\log (\rho{}L))$  & $L = \max w(e)$, $\log ({\rho{}L}) \leq \rho \leq n$\\
                     & Spencer and Shi~\cite{ShiS:jalg99} & $O(\tfrac{n^3}{\rho^2}\log n \log \tfrac{n}\rho + m \log n)$ or
$O((\tfrac{n^3}{\rho^3} + \tfrac{mn}{\rho}) \log n)$ & $O(\rho\log{n})$ \\
                     & Cohen~\cite{cohen1997using} & $O(n^2 + n^3/\rho^2)$ & $O(\rho\cdot{}\polylog(n))$\\
                                          & \emph{\textsf{This work}} & $O((m+n\rho)\log n)$ & $O(\tfrac{n}\rho\log n \log\rho L)$ &  \\
                     & & preproc: $O(m\log n+n\rho^2)$ or $O(m\log n+n\rho^2\log\rho)$ & $O(\rho^2)$ or $O(\rho\log \rho)$\\
    \bottomrule
  \end{tabular}\par
  \caption{Work/depth bounds for exact {\normalfont\sssp} algorithms that have subcubic work.}\vspace{-1.5em}
  \label{table:cost-comparison}
\end{table*}


There has been a number of research papers on the topic of parallel
single-source shortest paths.  In this section, we discuss prior works that are
most relevant to ours (for a more comprehensive survey, see~\cite{MS03} and the
references therein).

Some early algorithms achieve a high degree of parallism (polylogarithm depth)
but with substantially more work than Dijkstra's algorithm.
Using matrix multiplications over semirings, Han et~al.~\cite{han1992efficient}
gives an algorithm with $O(\log^2 n)$ depth and
$O(n^3 (\log \log n/\log n)^{1/3})$ work that, in fact, solves the
\emph{all-pairs} shortest-path problem.  With randomization, this algorithm can
be implemented in $O(\log n)$ depth and $O(n^3 \log n)$
work~\cite{frieze1984parallel}.

In another line of work, Driscoll et~al.~\cite{driscoll1988relaxed}, refining
the approach of Paige and Kruskal~\cite{PaigeK:icpp85}, present an
$O(n\log n)$-depth algorithm with Dijkstra's work bound.  Later, Brodal
et~al.~\cite{brodal1998parallel} improve the depth to $O(n)$; however, the
algorithm needs $O(m\log n)$ work.  We summarize in
Table~\ref{table:cost-comparison} the cost bounds for exact \sssp algorithms
that have subcubic work.

By allowing a slight increase in work in exchange for a better depth bound, many
algorithms have been proposed.  Ullman and Yannakakis~\cite{ullman1991high}
describe a randomized parallel breadth-first search (BFS) that solves unweighted
\sssp in $\otilde(m\sqrt{n} + nm/t + n^3/t^4)$ work and $\otilde(t)$
depth for a tuneable parameter $t \leq \sqrt{n}$.  The algorithm works by
performing limited searches from a number of locations and adding shortcut edges
with appropriate distances to speed up later traversal.  Klein and
Subramanian~\cite{klein1997randomized} extended this idea to weighted graphs,
resulting in an algorithm that runs in $O(\sqrt{n} \log K \log{n})$ depth and
$O(m\sqrt{n}\log K \log{n})$ work, where $K$ is the maximum weighted distance
from $s$ to any node reachable from $s$.  For undirected graphs,
Cohen~\cite{cohen1997using} presents an algorithm with
$O(\rho\cdot{}\polylog(n))$ depth and $O(n^2 + n^3/\rho^2)$ work. Shi and
Spencer~\cite{ShiS:jalg99} shows an algorithm with $O(\rho\log n)$ depth, and
$O((n^3/\rho^2)\log n \log(n/\rho) + m \log n)$ or
$O((n^3/\rho^3 + mn/\rho) \log n)$ work.

More recently, Meyers and Sanders~\cite{MS03} describe an algorithm called
$\Delta$-stepping and analyze it for various random graph
settings. 
Although the algorithm works well on general graphs in practice, no theoretical
guarantees are known.  In addition to these results, better algorithms have been
developed for special classes of graphs such as planar graphs
(e.g.,~\cite{traff1996simple,klein1993linear}) and separator-decomposable
graphs~\cite{Cohen:jalg96}.  There have also been approximation algorithms for
\sssp. Cohen~\cite{cohen2000polylog} describes a $(1+\varepsilon)$-algorithm for
undirected graphs that runs in $O(\polylog(n))$ depth and $O((m+n)n^{\alpha})$
work for $\alpha > 0$.  Using an alternative hopset construction of Miller
et~al.~\cite{MillerPVX15}, Cohen's algorithm can be made to run in
$O(m\cdot{}\polylog(n))$ work and $n^{1-\alpha}$ depth for $\alpha > 0$.


\section{Conclusion}
We presented a parallel algorithm for the single-source shortest-path problem
that after preprocessing, runs in $O(m\log n)$ work and
$O(\tfrac{n}{\rho} \log n \log {\rho{}L})$ depth, where $\rho$ is
a user-defined parameter. Compared to an optimal sequential implementation, this
is only a factor of $O(\log n)$ from being work-efficient.
Indeed, the algorithm offers the best-known tradeoffs between work and depth,
and is conceptually simpler than prior algorithms.  We leave open the question
of finding a globally-optimal way to add shortcut edges for $k > 1$.


\bibliographystyle{abbrv}
\bibliography{ref}
\clearpage
 \appendix
 \hide{\section{Analysis on the Work Bound}
\label{app:work}
In Section \ref{sec:workanddepth} we show the work of \AlgName{} can be dominated by $\sum_{i=1}^{t}{a_ik\log \frac{n}{a_i}}$, where $a_i=|A_i|$ is the number of vertices visited in $i$-th step, and $\sum_i a_i=m$.
Assume in total the algorithm runs $t$ steps, then $t=O(\frac{n}{\rho}\log \rho L)$\footnote{Notice that (also stated in Section \ref{sec:workanddepth}) throughout this section we assume $r(v)\le \rk{v}$ and $|B(v,r(v))|\ge \rho$ for each $v\in V$.}.

We bound the sum of $\sum_i{a_i k\log n}$ by two parts. For all $a_i\ge \rho$, the sum is:
\begin{align*}
\sum_{i=1}^{t}{a_ik\log \frac{n}{a_i}} &\le \sum_{i=1}^{t}a_i k\log \frac{n}{\rho}\\
&=km\log \frac{n}{\rho}
\end{align*}
For all $a_i<\rho$, the sum is:
\begin{align*}
\sum_{i=1}^{t}{a_i\cdot k\log \frac{n}{a_i}} &\le \sum_{i=1}^{t}\rho\cdot k\log n\\
&=k\rho\cdot\frac{n}{\rho}\log \rho\log n=kn\log n\log \rho
\end{align*}
Thus the total work can be bounded by $km\log \frac{n}{\rho}+kn\log n\log \rho$. Considering $\rho \le \sqrt{n}$, the bound is $O\left(k(m+n\log n)\log \frac{n}{\rho}\right)$.
}

\section{Tables for Experiments}
Tables \ref{tab:addingedgegreedy} to \ref{tab:stepreducedweighted} show additional experimental results. Tables \ref{tab:addingedgegreedy} and \ref{tab:addingedgedp} show the number of added edges by the greedy heuristic and the dynamic programming heuristic, respectively, varying both $k$ and $\rho$. Table \ref{tab:stepunweighted} shows the number of steps required by \AlgName{} on different unweighted graphs with different values of $\rho$, and Table \ref{tab:stepreducedunweighted} shows the corresponding reduced factor compared to when $\rho=1$ (standard BFS). Table \ref{tab:stepweighted} shows the number of steps required by \AlgName{} on different weighted graphs with different values of $\rho$, and Table \ref{tab:stepreducedweighted} shows the corresponding reduced factor compared to when $\rho=1$ (standard Dijkstra's).

\begin{table*}[h!]
  \centering\small
    \begin{tabular}{@{}r@{ }|@{}r@{ }r@{ }r@{ }r@{ }|@{}r@{ }|@{}r@{ }r@{ }r@{ }r@{ }|@{}r@{ }|@{}r@{ }r@{ }r@{ }r@{ }|@{}r@{}}
          & \multicolumn{5}{c|}{\textbf{Roadmap of Pennsylvania}}   & \multicolumn{5}{@{ }c@{ }|}{\textbf{Webgraph of Stanford}} & \multicolumn{5}{c}{\textbf{2D-grid}} \\
          & \multicolumn{5}{c|}{($|V|$=1.09M, $|E|$=3.08M)} & \multicolumn{5}{@{}c@{}|}{($|V|$=281k, $|E|$=3.98M)} & \multicolumn{5}{c}{($|V|$=1M, $|E|$=2M)} \\
          \hline
    & \multicolumn{4}{c|}{\textbf{k}}         & \multicolumn{1}{@{}c@{}|}{\textbf{red.}} & \multicolumn{4}{c|}{\textbf{k}}         & \multicolumn{1}{@{}c@{}|}{\textbf{red.}} & \multicolumn{4}{c|}{\textbf{k}}         & \multicolumn{1}{@{}c@{}}{\textbf{red.}} \\
          \cline{2-5}\cline{7-10}\cline{12-15}
          \multicolumn{1}{c|}{$\bm \rho$}& \multicolumn{1}{c}{\textbf{2}}     & \multicolumn{1}{c}{\textbf{3}}     & \multicolumn{1}{c}{\textbf{4}}     & \multicolumn{1}{c|}{\textbf{5}}     & \multicolumn{1}{@{}c@{}|}{\textbf{rounds}} & \multicolumn{1}{c}{\textbf{2}}     & \multicolumn{1}{c}{\textbf{3}}     & \multicolumn{1}{c}{\textbf{4}}     & \multicolumn{1}{c|}{\textbf{5}}    & \multicolumn{1}{@{}c@{}|}{\textbf{rounds}} & \multicolumn{1}{c}{\textbf{2}}     & \multicolumn{1}{c}{\textbf{3}}     & \multicolumn{1}{c}{\textbf{4}}     & \multicolumn{1}{c|}{\textbf{5}}     & \multicolumn{1}{@{}c@{}}{\textbf{rounds}} \\
          \hline
    \textbf{10}    & 1.67  & 0.41  & 0.05  & 0.01  & 3.00  & 3.11  & 0.02  & 0.01  & 0.00  & 3.48  & 0.36  & 0.00  & 0.00  & 0.00  & 3.00 \\
    \textbf{20}    & 3.79  & 2.38  & 0.84  & 0.23  & 3.88  & 9.91  & 3.06  & 0.09  & 0.01  & 5.03  & 5.75  & 0.46  & 0.00  & 0.00  & 4.00 \\
    \textbf{50}    & 10.34 & 6.05  & 5.65  & 3.71  & 5.04  & 47.57 & 10.74 & 3.40  & 0.13  & 7.71  & 16.05 & 8.40  & 9.54  & 0.67  & 6.01 \\
    \textbf{100}   & 20.33 & 13.64 & 8.85  & 8.16  & 6.13  & 109.98 & 39.99 & 20.96 & 8.73  & 10.25 & 29.59 & 22.02 & 10.52 & 11.43 & 8.02 \\
    \textbf{200}   & 39.92 & 26.35 & 20.15 & 14.51 & 7.85  & 188.92 & 67.25 & 45.54 & 17.96 & 12.72 & 48.40 & 41.34 & 28.03 & 12.73 & 11.04 \\
    \textbf{500}   & 97.58 & 64.72 & 48.49 & 37.64 & 10.76 & 337.34 & 141.58 & 119.03 & 63.69 & 14.92 & 126.09 & 99.22 & 55.62 & 64.75 & 17.12 \\
    \textbf{1000}  & 192.00 & 127.45 & 95.55 & 75.84 & 13.74 & 529.14 & 208.66 & 219.21 & 149.20 & 15.17 & 243.12 & 181.50 & 129.26 & 108.37 & 23.18 \\
    \end{tabular}%
  \caption{Factors of additional edges added using greedy heuristic.}
  \label{tab:addingedgegreedy}%
\end{table*}%

\begin{table*}[h!]
 \centering\small
    \begin{tabular}{@{}r@{ }|@{}r@{ }r@{ }r@{ }r@{ }|@{}r@{ }|rrrr|r|@{}r@{ }r@{ }r@{ }r@{ }|@{}r@{}}
          & \multicolumn{5}{c|}{\textbf{Roadmap of Pennsylvania}}   & \multicolumn{5}{@{ }c@{ }|}{\textbf{Webgraph of Stanford}} & \multicolumn{5}{c}{\textbf{2D-grid}} \\
          & \multicolumn{5}{c|}{($|V|$=1.09M, $|E|$=3.08M)} & \multicolumn{5}{@{}c@{}|}{($|V|$=281k, $|E|$=3.98M)} & \multicolumn{5}{c}{($|V|$=1M, $|E|$=2M)} \\
          \hline
    & \multicolumn{4}{c|}{\textbf{k}}         & \multicolumn{1}{@{}c@{}|}{\textbf{red.}} & \multicolumn{4}{c|}{\textbf{k}}         & \multicolumn{1}{@{}c@{}|}{\textbf{red.}} & \multicolumn{4}{c|}{\textbf{k}}         & \multicolumn{1}{@{}c@{}}{\textbf{red.}} \\
          \cline{2-5}\cline{7-10}\cline{12-15}
          \multicolumn{1}{c|}{$\bm \rho$}& \multicolumn{1}{c}{\textbf{2}}     & \multicolumn{1}{c}{\textbf{3}}     & \multicolumn{1}{c}{\textbf{4}}     & \multicolumn{1}{c|}{\textbf{5}}     & \multicolumn{1}{@{}c@{}|}{\textbf{rounds}} & \multicolumn{1}{c}{\textbf{2}}     & \multicolumn{1}{c}{\textbf{3}}     & \multicolumn{1}{c}{\textbf{4}}     & \multicolumn{1}{c|}{\textbf{5}}    & \multicolumn{1}{@{}c@{}|}{\textbf{rounds}} & \multicolumn{1}{c}{\textbf{2}}     & \multicolumn{1}{c}{\textbf{3}}     & \multicolumn{1}{c}{\textbf{4}}     & \multicolumn{1}{c|}{\textbf{5}}     & \multicolumn{1}{@{}c@{}}{\textbf{rounds}} \\
          \hline
    \textbf{10}    & 0.95  & 0.12  & 0.01  & 0.00  & 3.00  & 0.02  & 0.01  & 0.01  & 0.00  & 3.48  & 0.25  & 0.00  & 0.00  & 0.00  & 3.00 \\
    \textbf{20}    & 2.70  & 0.90  & 0.18  & 0.04  & 3.88  & 0.05  & 0.02  & 0.01  & 0.01  & 5.03  & 3.95  & 0.25  & 0.00  & 0.00  & 4.00 \\
    \textbf{50}    & 7.78  & 3.59  & 1.89  & 0.72  & 5.04  & 0.20  & 0.06  & 0.04  & 0.03  & 7.71  & 12.16 & 6.21  & 4.06  & 0.36  & 6.01 \\
    \textbf{100}   & 16.09 & 8.09  & 4.40  & 2.58  & 6.13  & 0.51  & 0.13  & 0.08  & 0.06  & 10.25 & 24.22 & 14.27 & 8.32  & 6.06  & 8.02 \\
    \textbf{200}   & 32.60 & 17.04 & 9.89  & 6.03  & 7.85  & 0.99  & 0.25  & 0.15  & 0.11  & 12.72 & 48.35 & 30.23 & 20.28 & 12.45 & 11.04 \\
    \textbf{500}   & 81.75 & 44.14 & 26.65 & 17.11 & 10.76 & 2.18  & 0.50  & 0.30  & 0.22  & 14.92 & 125.96 & 80.09 & 54.44 & 42.26 & 17.12 \\
    \textbf{1000}  & 162.91 & 89.30 & 54.82 & 35.95 & 13.74 & 3.92  & 0.66  & 0.34  & 0.24  & 15.17 & 241.30 & 154.97 & 110.87 & 84.87 & 23.18 \\
    \end{tabular}%
  \caption{Factors of additional edges added using DP heuristic.}
  \label{tab:addingedgedp}%
\end{table*}%

\begin{table*}[!h]
  \centering
    \begin{tabular}{r|c|>{\centering}p{1.5cm}>{\centering}p{1.5cm}|>{\centering}p{1.8cm}>{\centering}p{1.8cm}|>{\centering}p{1.5cm}p{1.5cm}<{\centering}}
    \multicolumn{2}{c|}{} & \multicolumn{2}{c|}{\textbf{Roadmaps}} & \multicolumn{2}{c|}{\textbf{Webgraphs}} & \multicolumn{2}{c}{\textbf{Grids}} \\
 \hline
    \multicolumn{2}{c|}{} & \textbf{Penn}  & \textbf{Texas} & \textbf{NotreDame} & \textbf{Stanford} & \textbf{2D}    & \textbf{3D} \\
    \hline
    \multicolumn{2}{c|}{\textbf{vertices}} & 1.09M & 1.39M & 325k & 281k & 1M & 1M \\
    \multicolumn{2}{c|}{\textbf{edges}} & 3.08M & 3.84M & 2.20M & 3.98M & 2M & 5.94M \\
    \hline
    \multirow{13}{*}{$\bm{\rho}$} & 1     & 619.12 & 761.06 & 28.09 & 108.92 & 1504.0  & 223.50 \\
          & \textbf{2}     & 309.32 & 380.31 & 13.77 & 54.23 & 751.76 & 111.50 \\
          & \textbf{5}     & 308.47 & 379.34 & 13.44 & 43.27 & 751.74 & 111.50 \\
          & \textbf{10}    & 206.30 & 253.71 & 13.32 & 31.29 & 501.14 & 74.50 \\
          & \textbf{20}    & 165.73 & 196.30 & 13.17 & 21.67 & 375.62 & 74.48 \\
          & \textbf{50}    & 123.01 & 151.13 & 12.38 & 14.13 & 250.32 & 55.48 \\
          & \textbf{100}   & 101.41 & 124.07 & 9.78  & 10.63 & 187.46 & 44.08 \\
          & \textbf{200}   & 78.61 & 96.92 & 8.47  & 8.56  & 136.24 & 36.48 \\
          & \textbf{500}   & 58.44 & 70.75 & 6.63  & 7.30  & 87.86 & 27.36 \\
          & \textbf{1000}  & 45.95 & 55.39 & 5.69  & 7.18  & 64.88 & 21.74 \\
          & \textbf{2000}  & 35.66 & 42.58 & 5.27  & 6.72  & 44.82 & 17.94 \\
          & \textbf{5000}  & 24.95 & 29.17 & 4.14  & 5.84  & 28.82 & 12.50 \\
          & \textbf{10000} & 18.54 & 21.33 & 3.83  & 5.76  & 20.18 & 10.00 \\
    \end{tabular}%
  \caption{Average number of rounds with different of $\rho$ on different unweighted graphs.}
  \label{tab:stepunweighted}%
\end{table*}%

\begin{table*}[!h]
  \centering
    \begin{tabular}{r|c|>{\centering}p{1.5cm}>{\centering}p{1.5cm}|>{\centering}p{1.8cm}>{\centering}p{1.8cm}|>{\centering}p{1.5cm}p{1.5cm}<{\centering}}
    \multicolumn{2}{c|}{} & \multicolumn{2}{c|}{\textbf{Roadmaps}} & \multicolumn{2}{c|}{\textbf{Webgraphs}} & \multicolumn{2}{c}{\textbf{Grids}} \\
 \hline
    \multicolumn{2}{c|}{} & \textbf{Penn}  & \textbf{Texas} & \textbf{NotreDame} & \textbf{Stanford} & \textbf{2D}    & \textbf{3D} \\
    \hline
    \multicolumn{2}{c|}{\textbf{vertices}} & 1.09M & 1.39M & 325k & 281k & 1M & 1M \\
    \multicolumn{2}{c|}{\textbf{edges}} & 3.08M & 3.84M & 2.20M & 3.98M & 2M & 5.94M \\
    \hline
    \multirow{13}{*}{$\bm{\rho}$}
          & \textbf{2}     & 2.00  & 2.00  & 2.04  & 2.01  & 2.00  & 2.00 \\
          & \textbf{5}     & 2.01  & 2.01  & 2.09  & 2.52  & 2.00  & 2.00 \\
          & \textbf{10}    & 3.00  & 3.00  & 2.11  & 3.48  & 3.00  & 3.00 \\
          & \textbf{20}    & 3.74  & 3.88  & 2.13  & 5.03  & 4.00  & 3.00 \\
          & \textbf{50}    & 5.03  & 5.04  & 2.27  & 7.71  & 6.01  & 4.03 \\
          & \textbf{100}   & 6.11  & 6.13  & 2.87  & 10.25 & 8.02  & 5.07 \\
          & \textbf{200}   & 7.88  & 7.85  & 3.32  & 12.72 & 11.04 & 6.13 \\
          & \textbf{500}   & 10.59 & 10.76 & 4.24  & 14.92 & 17.12 & 8.17 \\
          & \textbf{1000}  & 13.47 & 13.74 & 4.94  & 15.17 & 23.18 & 10.28 \\
          & \textbf{2000}  & 17.36 & 17.87 & 5.33  & 16.21 & 33.56 & 12.46 \\
          & \textbf{5000}  & 24.81 & 26.09 & 6.79  & 18.65 & 52.19 & 17.88 \\
          & \textbf{10000} & 33.39 & 35.68 & 7.33  & 18.91 & 74.53 & 22.35 \\

    \end{tabular}%
  \caption{The reduced factor of rounds on unweighted graphs comparing to BFS rounds.}
  \label{tab:stepreducedunweighted}%
\end{table*}%

\begin{table*}[!h]
  \centering
    \begin{tabular}{r|c|>{\centering}p{1.5cm}>{\centering}p{1.5cm}|>{\centering}p{1.8cm}>{\centering}p{1.8cm}|>{\centering}p{1.5cm}p{1.5cm}<{\centering}}
    \multicolumn{2}{c|}{} & \multicolumn{2}{c|}{\textbf{Roadmaps}} & \multicolumn{2}{c|}{\textbf{Webgraphs}} & \multicolumn{2}{c}{\textbf{Grids}} \\
 \hline
    \multicolumn{2}{c|}{} & \textbf{Penn}  & \textbf{Texas} & \textbf{NotreDame} & \textbf{Stanford} & \textbf{2D}    & \textbf{3D} \\
    \hline
    \multicolumn{2}{c|}{\textbf{vertices}} & 1.09M & 1.39M & 325k & 281k & 1M & 1M \\
    \multicolumn{2}{c|}{\textbf{edges}} & 3.08M & 3.84M & 2.20M & 3.98M & 2M & 5.94M \\
    \hline
    \multirow{10}{*}{$\bm{\rho}$}
           &          \textbf{1} &       986K &      1252K &      35.6K &      30.0K &       965K &       239K \\

           &          \textbf{2} &    26479.9 &    34673.4 &     1953.7 &     2203.3 &    33592.2 &    11046.1 \\

           &          \textbf{5} &     2294.5 &     3123.5 &      571.3 &      759.2 &     3495.8 &      722.4 \\

           &         \textbf{10} &      872.6 &     1206.5 &      387.2 &      562.3 &     1385.0 &      261.9 \\

           &         \textbf{20} &      455.0 &      634.1 &      274.9 &      432.2 &      722.9 &      137.8 \\

           &         \textbf{50} &      245.0 &      343.0 &      174.6 &      293.7 &      375.1 &       76.1 \\

           &        \textbf{100} &      167.2 &      233.7 &      118.8 &      219.3 &      246.9 &       54.1 \\

           &        \textbf{200} &      119.8 &      166.9 &       83.7 &      166.0 &      166.9 &       40.2 \\

           &        \textbf{500} &       81.1 &      111.3 &       58.4 &      120.0 &      102.1 &       28.1 \\

           &       \textbf{1000} &       61.1 &       83.2 &       45.0 &       93.6 &       71.1 &       21.7 \\

    \end{tabular}%
  \caption{Average number of rounds with different of $\rho$ on different weighted graphs.}
  \label{tab:stepweighted}%
\end{table*}%

\begin{table*}[!h]
  \centering
    \begin{tabular}{r|c|>{\centering}p{1.5cm}>{\centering}p{1.5cm}|>{\centering}p{1.8cm}>{\centering}p{1.8cm}|>{\centering}p{1.5cm}p{1.5cm}<{\centering}}
    \multicolumn{2}{c|}{} & \multicolumn{2}{c|}{\textbf{Roadmaps}} & \multicolumn{2}{c|}{\textbf{Webgraphs}} & \multicolumn{2}{c}{\textbf{Grids}} \\
 \hline
    \multicolumn{2}{c|}{} & \textbf{Penn}  & \textbf{Texas} & \textbf{NotreDame} & \textbf{Stanford} & \textbf{2D}    & \textbf{3D} \\
    \hline
    \multicolumn{2}{c|}{\textbf{vertices}} & 1.09M & 1.39M & 325k & 281k & 1M & 1M \\
    \multicolumn{2}{c|}{\textbf{edges}} & 3.08M & 3.84M & 2.20M & 3.98M & 2M & 5.94M \\
    \hline
    \multirow{9}{*}{$\bm{\rho}$}
           &          \textbf{2} &       37.2 &       36.1 &       18.2 &       13.6 &       28.7 &       21.6 \\

           &          \textbf{5} &      429.7 &      400.8 &       62.3 &       39.6 &      276.1 &      330.6 \\

           &         \textbf{10} &     1130.0 &     1037.7 &       92.0 &       53.4 &      696.8 &      912.0 \\

           &         \textbf{20} &     2167.0 &     1974.5 &      129.5 &       69.5 &     1334.9 &     1733.3 \\

           &         \textbf{50} &     4024.5 &     3650.1 &      203.9 &      102.3 &     2572.7 &     3138.5 \\

           &        \textbf{100} &     5897.1 &     5357.3 &      299.7 &      137.0 &     3908.6 &     4414.9 \\

           &        \textbf{200} &     8230.4 &     7501.5 &      425.4 &      180.9 &     5782.0 &     5941.4 \\

           &        \textbf{500} &    12157.8 &    11248.9 &      609.7 &      250.3 &     9451.7 &     8499.8 \\

           &       \textbf{1000} &    16137.5 &    15048.1 &      791.2 &   320.9 &    13572.8 &    11006.6 \\

    \end{tabular}%
  \caption{The reduced factor of rounds on weighted graphs comparing to $\rho=1$.}
  \label{tab:stepreducedweighted}%
\end{table*}%

\end{document}